\documentclass[12pt]{article}
\usepackage{amsmath,amsthm,amssymb,epsfig,euscript,array,cancel,color}
\usepackage[colorlinks=true]{hyperref}
\usepackage{verbatim}
\usepackage{url}
\usepackage{bbm}

\usepackage[T1]{fontenc}    
\usepackage[utf8]{inputenc}

\usepackage[paper=a4paper,dvips,top=2.54cm,left=2.74cm,right=2.34cm,
    foot=1cm,bottom=2.54cm]{geometry}

\theoremstyle{definition} 

\theoremstyle{definition}

\numberwithin{trial}{subsection}

\theoremstyle{remark}

\newcounter{multieqs}

\newcommand{\be}{\begin{equation}}
\newcommand{\ee}{\end{equation}}
\newcommand{\eq}[1]{(\ref{#1})}
\newcommand{\bit}{\begin{itemize}}  \newcommand{\eit}{\end{itemize}}

\newcommand{\bm}[1]{\mbox{\boldmath $#1$}}

\def\bd{\begin{document}}
\def\ed{\end{document}}
\def\nn{\nonumber}
\def\bea{\begin{eqnarray}}
\def\eea{\end{eqnarray}}
\let\bm=\bibitem

\def\la{\langle}
\def\ra{\rangle}

\def\npb#1#2#3{Nucl. Phys. {\bf{B#1}} #3 (#2)}
\def\plb#1#2#3{Phys. Lett. {\bf{#1B}} #3 (#2)}
\def\prl#1#2#3{Phys. Rev. Lett. {\bf{#1}} #3 (#2)}
\def\prd#1#2#3{Phys. Rev. {D \bf{#1}} #3 (#2)}
\def\cmp#1#2#3{Comm. Math. Phys. {\bf{#1}} #3 (#2)}
\def\cqg#1#2#3{Class. Quantum Grav. {\bf{#1}} #3 (#2)}
\def\nppsa#1#2#3{Nucl. Phys. B (Proc. Suppl.) {\bf{#1A}}#3 (#2)}
\def\ap#1#2#3{Ann. of Phys. {\bf{#1}} #3 (#2)}
\def\ijmp#1#2#3{Int. J. Mod. Phys. {\bf{A#1}} #3 (#2)}
\def\rmp#1#2#3{Rev. Mod. Phys. {\bf{#1}} #3 (#2)}
\def\mpla#1#2#3{Mod. Phys. Lett. {\bf A#1} #3 (#2)}
\def\jhep#1#2#3{J. High Energy Phys. {\bf #1} #3 (#2)}
\def\atmp#1#2#3{Adv. Theor. Math. Phys. {\bf #1} #3 (#2)}

\def\N{{\cal N}}
\def\sst{\scriptscriptstyle}
\def\thetabar{\bar\theta}
\def\Tr{{\rm Tr}}
\def\one{\mbox{1 \kern-.59em {\rm l}}}

\def\a{\alpha}      \def\da{{\dot\alpha}}  \def\dA{{\dot A}}
\def\b{\beta}       \def\db{{\dot\beta}}  
\def\g{\gamma}  \def\G{\Gamma}  \def\dc{{\dot\gamma}}  
\def\d{\delta}  \def\D{\Delta}  \def\ddt{\dot\delta}  
\def\e{\epsilon}        \def\ve{\varepsilon}  
\def\f{\phi}    \def\F{\Phi}    \def\vvf{\f}  
\def\h{\eta}  
\def\k{\kappa}  
\def\l{{\lambda}} \def\L{\Lambda}  
\def\m{\mu} \def\n{\nu}  
\def\o{\omega}  
\def\p{\pi} \def\P{\Pi}  
\def\r{\rho}  
\def\s{\sigma}  \def\S{\Sigma}  
\def\t{\tau}  
\def\th{\theta} \def\Th{\Theta} \def\vth{\vartheta}  
\def\X{\Xeta}  
\def\z{\zeta}  

\def\na{\nabla}  

\def\cA{{\cal A}} \def\cB{{\cal B}} \def\cC{{\cal C}}  
\def\cD{{\cal D}} \def\cE{{\cal E}} \def\cF{{\cal F}}  
\def\cG{{\cal G}} \def\cH{{\cal H}} \def\cI{{\cal I}}  
\def\cJ{{\cal J}} \def\cK{{\cal K}} \def\cL{{\cal L}}  
\def\cM{{\cal M}} \def\cN{{\cal N}} \def\cO{{\cal O}}  
\def\cP{{\cal P}} \def\cQ{{\cal Q}} \def\cR{{\cal R}}  
\def\cS{{\cal S}} \def\cT{{\cal T}} \def\cU{{\cal U}}  
\def\cV{{\cal V}} \def\cW{{\cal W}} \def\cX{{\cal X}}  
\def\cY{{\cal Y}} \def\cZ{{\cal Z}}

\def\ua{{\underline{\alpha}}} 
 \def\ub{\underline{\phantom{\alpha}}\!\!\!\beta}  
\def\uc{\underline{\phantom{\alpha}}\!\!\!\gamma}  
\def\um {{\underline{\mu}}}  
\def\ud{{\underline{\delta}}} 
\def\ue{\underline\epsilon}  
\def\una{{\underline a}}\def\uA{{\underline A}}  
\def\unb{{\underline b}}\def\uB{{\underline B}} 
\def\unc{{\underline c}}\def\uC{{\underline C}}  
\def\und{{\underline d}}\def\uD{{\underline D}}  
\def\une{{\underline e}}\def\uE{{\underline E}}  
\def\unf{{\underline{\phantom{e}}\!\!\!\! f}}\def\uF{{\underline F}}  
\def\unm{{\underline m}\def\uM{\underline M}} 
\def\unn{{\underline n}\def\uN{\underline N}} 
\def\unp{{\underline{\phantom{a}}\!\!\! p}}\def\uP{{\underline P}}  
\def\unq{{\underline{\phantom{a}}\!\!\! q}}  
\def\uQ{{\underline{\phantom{A}}\!\!\!\! Q}}  
\def\uH{{\underline{H}}}  
\def\uM{{\underline{M}}}
\def\uN{{\underline{N}}}
\def\unl{{\underline{l}}}
  
\def\As {{A \hspace{-6.4pt} \slash}\;}  
\def\bs {{b \hspace{-6.4pt} \slash}\;}  
\def\Ds {{D \hspace{-6.4pt} \slash}\;}
\def\Gts {{\Gt \hspace{-6.4pt} \slash}\;}
\def\ds {{\del \hspace{-6.4pt} \slash}\;}  
\def\ss {{\s \hspace{-6.4pt} \slash}\;}  
\def\ks {{ k \hspace{-6.4pt} \slash}\;}  
\def\ps {{p \hspace{-6.4pt} \slash}\;}   
\def\xs {{x \hspace{-6.4pt} \slash}\;}  
\def\pas {{{p_1} \hspace{-6.4pt} \slash}\;}  
\def\pbs {{{p_2} \hspace{-6.4pt} \slash}\;}   
\def\cFs {{{\cal F} \hspace{-6.4pt} \slash}\;}

\def\Ah{{\hat{A}}}  
\def\Dh{{\hat{D}}}
\def\Gh{{\hat{G}}}
\def\Fh{{\hat{F}}}
\def\Ih{{\hat{I}}} 
\def\Jh{{\hat{J}}} 
\def\Kh{{\hat{K}}}
\def\Lh{{\hat{L}}} 
\def\Ph{{\hat{P}}}
\def\Rh{{\hat{R}}}
\def\Vh{{\hat{V}}} 
\def\Xh{{\hat{X}}}
 
\def\ah{{\hat{a}}}
\def\bh{{\hat{b}}}
\def\ch{{\hat{c}}}
\def\gh{{\hat{g}}}
\def\dh{{\hat{d}}}
\def\hh{{\hat{h}}}
\def\uh{{\hat{u}}}  
\def\vh{{\hat{v}}}
\def\xh{{\hat{x}}} 
\def\yh{{\hat{y}}}
\def\zh{{\hat{z}}}
\def\ph{{\hat{p}}}
\def\qh{{\hat{q}}}
\def\thh{{\hat{t}}}  
\def\xih{\hat{\xi}}  
\def\Psih{\hat{\Psi}}    
\def\mh{{\hat{m}}}
\def\nh{{\hat{n}}}
\def\ih{{\hat{i}}}
\def\jh{{\hat{j}}}
\def\kh{{\hat{k}}}
\def\aah{{\hat{\alpha}}}
\def\bbh{{\hat{\beta}}}
\def\ggh{{\hat{\gamma}}}
\def\llh{{\hat{\ell}}} 
\def\ph{{\hat{p}}}

\def\psit{\tilde{\psi}}  
\def\Psit{\tilde{\Psi}}   
\def\Psibt{\tilde{\bar{Psi}}}  

\def\st{\tilde{\sigma}}  

\def\delt{\tilde{\delta}}
\def\Phit{\tilde{\Phi}}   
\def\Phitb{\overline{\tilde{Phi}}}  
\def\tht{\tilde{\th}}  
\def\lt{\tilde{\l}}
\def\chit{\tilde{\chi}}   
\def\phit{\tilde{\phi}} 

\def\At{\tilde{A}}
\def\Bt{\tilde{B}}
\def\Ct{\tilde{C}}
\def\Dt{\tilde{D}}
\def\Et{\tilde{E}}
\def\Ft{\tilde{F}}
\def\Gt{\tilde{G}}
\def\Ht{\tilde{H}}
\def\It{\tilde{I}}
\def\Jt{\tilde{J}}
\def\Qt{\tilde{Q}}  
\def\Rt{\tilde{R}}  
\def\Mt{\tilde{M }}  
\def\Nt{\tilde{N}}   
\def\St{\tilde{S}}
\def\Vt{\tilde{V}}
\def\Xt{\tilde{X}} 
\def\at{\tilde{a}}
\def\ct{\tilde{c}}
\def\dt{\tilde{d}}
\def\htt{\tilde{h}} 
\def\ft{\tilde{f}}
\def\gt{\tilde{g}}
\def\pt{\tilde{p}}  
\def\qt{\tilde{q}}  
\def\vt{\tilde{v}}  
\def\nt{\tilde{n}}  
\def\ut{\tilde{u}}  
\def\wt{\tilde{w}}  
\def\zt{\tilde{z}} 
\def\xt{\tilde{x}} 
\def\yt{\tilde{y}} 
\def\Psit{\tilde{\Psi}}
\def\vphit{\tilde{\varphi}}

\def\gamt{\tilde{\gamma}}

\def\Tt{\tilde{T}}

\def\eb{\bar{\epsilon}} 
\def\delb{\bar{\partial}}  
\def\thb{\bar{\theta}}
\def\Thb{{\bar{\Theta}}}
\def\mub{\bar{\mu}}
\def\lamb{\bar{\l}}
\def\psib{\bar{\psi}}
\def\sb{\bar{\sigma}}
\def\xib{\bar{\xi}}
\def\chib{\bar{\chi}}

\def\Psib{\bar{\Psi}}
\def\Phib{\bar{\Phi}}
\def\Lamb{\bar{\Lambda}}
\def\Sb{{\overline \Sigma}}
\def\cb{\bar{c}}
\def\hb{\bar{h}}
\def\qb{\bar{q}}
\def\wb{\bar{w}}
\def\zb{{\bar{z}}}
\def\Hb{\bar{H}}
\def\Qb{{\bar Q}}
\def\Omegab{\overline{\Omega}}
\def\ob{\overline{\omega}}
\def\Gab{{\bar{\Gamma}}}

\def\Ab{{\overline A}} \def\Bb{{\overline B}} \def\Cb{{\overline C}}  
\def\Db{{\overline D}} \def\Eb{{\overline E}} \def\Fb{{\overline F}}  
\def\Gb{{\overline G}} 
\def\Ib{{\overline I}}  
\def\Jb{{\overline J}} \def\Kb{{\overline K}} \def\Lb{{\overline L}}  
\def\Mb{{\overline M}} \def\Nb{{\overline N}} \def\Ob{{\overline O}}  
\def\Pb{{\overline P}}  \def\Rb{{\overline R}}  
 \def\Tb{{\overline T}} \def\Ub{{\overline U}}  
\def\Vb{{\overline V}} \def\Wb{{\overline W}} \def\Xb{{\overline X}}  
\def\Yb{{\overline Y}} \def\Zb{{\overline Z}}  

\def\fb{{\overline f}}
\def\gb{{\overline g}}
\def\mb{{\overline m}}
\def\lb{{\overline l}}
\def\yb{{\overline y}}
  
\def\ldel{{\overleftarrow{\del}}}
\def\rdel{{\overrightarrow{\del}}}
\def\ldeldel{{\overleftarrow{\del^2}}}
\def\rdeldel{{\overrightarrow{\del^2}}}
\def\ldelb{{\overleftarrow{\bar{\del}}}}
\def\rdelb{{\overrightarrow{\bar{\del}}}}

\def\ba{{\bf a}} 
\def\bk{{\bf k}}  
\def\bl{{\bf l}}  
\def\bp{{\bf p}}  
\def\bq{{\bf q}}  
\def\br{{\bf r}}
\def\bt{{\bf t}}
\def\bu{{\bf u}}
\def\bv{{\bf v}}
\def\bx{{\bf x}}  
\def\by{{\bf y}}  
\def\bR{{\bf R}}  
\def\bV{{\bf V}}

\def\bone{{\bf 1}}  

\def\va{{\vec a}}
\def\vk{{\vec k}}
\def\vp{{\vec p}}
\def\vq{{\vec q}}
\def\vx{{\vec x}}
\def\vy{{\vec y}}
\def\vu{{\vec u}}
\def\vv{{\vec v}}

\def\vs{{\vec \sigma}}
\def\vtau{{\vec \tau}}

\newcommand{\ov}[1]{\overrightarrow{#1}}

\def\frA{\mathfrak{A}}
\def\frB{\mathfrak{B}}
\def\frC{\mathfrak{C}}
\def\frD{\mathfrak{D}}
\def\frE{\mathfrak{E}}
\def\frF{\mathfrak{F}}
\def\frG{\mathfrak{G}}
\def\frH{\mathfrak{H}}
\def\frM{\mathfrak{M}}
\def\frN{\mathfrak{N}}
\def\frR{\mathfrak{R}}
\def\frW{\mathfrak{W}}

\def\fra{\mathfrak{a}}
\def\frb{\mathfrak{b}}
\def\frf{\mathfrak{f}}
\def\frg{\mathfrak{g}}
\def\frh{\mathfrak{h}}
\def\frl{\mathfrak{l}}
\def\frs{\mathfrak{s}}
\def\fri{\mathfrak{i}}
\def\frj{\mathfrak{j}}

\def\ma{\mathfrak{a}}
\def\mg{\mathfrak{g}}
\def\mR{\mathfrak{R}}
\def\mN{\mathfrak{N}}

\def\d{\delta}\def\D{\Delta}\def\ddt{\dot\delta}  
  
\def\pa{\partial} \def\del{\partial}  
\def\xx{\times}  
\def\uno{\mbox{1 \kern-.59em {\rm l}}}    
  
\def\trp{^{\top}}  
\def\inv{^{-1}}  
\def\dag{{^{\dagger}}}  
\def\pr{^{\prime}}  
  
\def\rar{\rightarrow}  
\def\lar{\leftarrow}  
\def\lrar{\leftrightarrow}  
  
\newcommand{\0}{\,\!}      
\def\one{1\!\!1\,\,}  
\def\im{\imath}  
\def\jm{\jmath}  
  
\newcommand{\tr}{\mbox{tr}}  
\newcommand{\slsh}[1]{/ \!\!\!\! #1}  
  
\def\vac{|0\rangle}  
\def\lvac{\langle 0|}  
  
\def\hlf{\frac{1}{2}}  
\def\ove#1{\frac{1}{#1}}

\newcommand{\ex}[1]{{\rm e}^{#1}} \def\ii{{\rm i}}  

\newcommand{\lrbrk}[1]{\left(#1\right)}
\newcommand{\lrsbrk}[1]{\left[#1\right]}
\newcommand{\lrcbrk}[1]{\left\{#1\right\}}
\newcommand{\sfrac}[2]{{\textstyle\frac{#1}{#2}}}

\def \thbb{\overline{\th\th}}
\newcommand \ol{\overline}
\def \lamb{\bar{\lambda}}
\def \vphi{\varphi}
\def \lambh{\hat{\bar{\lambda}}}
\def \lh{\hat{\lambda}}
\def \dd{\ddagger}

\def\eh{{\hat{e}}}
\def\fh{{\hat{f}}}
\def\lh{{\hat{l}}}
\def\rh{{\hat{r}}}
\def\wh{{\hat{w}}}
\renewcommand{\mh}{{\hat{m}}}

\def \DBI{{\text{DBI}}}
\def\et{{\tilde{\e}}}
\def\w{{\wedge}}
\def\bbV{{\mathbb{V}}}
\def\M{{(\text{M})}}
\def\T{{(\text{T})}}

\def\Hbt{{\tilde{\bar{H}}}}
\def\Fbt{{\tilde{\bar{F}}}}

\def\rb{{\bar{r}}}

\newcommand{\PB}[2]{\{#1,#2\}}
\newcommand{\DB}[2]{[#1,#2]_{\textrm{D}}}
\newcommand{\FJB}[2]{[#1,#2]_{\textrm{FJ}}}

\author{Jarunee Sanongkhun\footnote{jaruneen57@email.nu.ac.th}~
and Pichet Vanichchapongjaroen\footnote{pichetv@nu.ac.th}~
\\
\\
{\small  
	\it The Institute for Fundamental Study ``The Tah Poe Academia Institute",}
\\
{\small\it Phitsanulok-Nakhonsawan Road, Naresuan University, Phitsanulok 65000, Thailand}
}

\title{\bf On constrained analysis and diffeomorphism invariance of generalised Proca theories}

\begin{document}
\maketitle

\abstract{
	In this paper we consider generalised Proca theories
	coupled to any background field
	and with time-time and time-space components
	of Hessian of the vector sector
	are zero, whereas the space-space part
	is non-degenerate.
	By using Faddeev-Jackiw analysis, we derive the conditions 
	that these theories have to satisfy in order
	for the vector sector to have three propagating degrees of freedom.
	Most of these conditions are trivialised due to 
	diffeomorphism invariance requirements.
	This leaves only a condition that
	a complicated combination of terms
	should not be trivially zero.
	This condition is therefore easy to be fulfilled.
	For completeness, we have also investigated
	on how diffeomorphism invariance helps
	in simplifying Faddeev-Jackiw brackets.}

\section{Introduction}	
	General Relativity has been a successful theory giving predictions
	which accurately agree with observations \cite{Will:2014kxa}.
	There are, however, results which cannot be described
	by General Relativity.
	One of these is the late-time accelerated expansion of the universe \cite{Perlmutter:1998np},\cite{Riess:1998cb}.
	As an attempt towards describing the mechanism behind this,
	one may consider modifying General Relativity.
	A simple way is to introduce a scalar-tensor theory,
	in which there is a scalar field introduced into the Lagrangian.
	By demanding that the extra scalar field
	has only one degree of freedom, thus is free from Ostrogradsky instabilities \cite{Ostrogradsky:1850fid},
	up to the second order derivative of this extra scalar field
	could appear in the Lagrangian.
	The scalar sector in flat spacetime
	is a Galilean theory \cite{Nicolis:2008in}.
	When generalised to curved spacetime
	and include the gravity sector,
	the resulting theory whose
	equations of motion for both scalar and gravity
	are of at most second order derivative
	can be constructed and is known as the Horndeski theory \cite{Horndeski:1974wa}, \cite{Deffayet:2009wt}, \cite{Deffayet:2009mn}, \cite{Deffayet:2011gz}.
	Further extensions to the Horndeski theory can actually be made \cite{Zumalacarregui:2013pma}, \cite{Gleyzes:2014dya}, \cite{Langlois:2015cwa}.
	For a review see \cite{Kobayashi:2019hrl}.
	
	An alternative direction is to consider a vector-tensor theory
	with Galilean-like interactions.
	It is pointed out by \cite{Deffayet:2013tca} that
	there is a no-go theorem preventing
	the inclusion of Galilean-like interactions
	in the case where the vector is massless.
	One then needs to turn to generalising the Proca theory,
	which is a massive vector theory.

	A generalised Proca theory describes
	a system of gauge field with derivative self-interaction.
	An original construction of generalised Proca theories
	is given in \cite{Tasinato:2014eka}, \cite{Heisenberg:2014rta}.
	The idea is to start from
	a form
	of Lagrangian of gauge field in flat spacetime
	with several constants to be determined.
	Demanding that the Hessian determinant vanishes
	ensures that the theory has constraints,
	and hence at most three propagating degrees of freedom.
	This requirement gives rise to conditions which relate some
	of the constants.
	After the theory in flat spacetime is constructed,
	an insight from Horndeski theory is made use
	to extend the theory to curved spacetime.

	The construction is confirmed and extended by \cite{Jimenez:2016isa},
	which systematically constructs the derivative self-interactions
	for the generalised Proca action beyond decoupling limit
	by using antisymmetric properties of Levi-Civita tensors.
	This finally gives rise to the Lagrangian of the form
	\be\label{LgenProca-curved}
	{\cal L}_{gen.Proca}=-\sqrt{-g}\frac{1}{4} F_{\m\n}F^{\m\n}+\sqrt{-g}\sum_{n=2}^{6} \b_{n}{\cal L}_n,
	\ee
	where
	$\b_{n}, n=2,3,4,5,6$ are arbitrary constants and ${\cal L}_n, n=2,3,4,5,6$
	are self-interactions of the gauge field given by
	\be\label{L2toL6genProca}
	\begin{split}
	{\cal L}_{2}&=G_2(A_\m, F_{\m\n}, \Ft_{\m\n}),\\
	{\cal L}_{3}&=G_3(Y)\na\cdot A,\\
	{\cal L}_{4}&=G_4(Y)R+G'_{4}(Y)[(\na\cdot A)^{2} - \na_{\r} A_{\s}\na^{\s} A^{\r}],\\
	{\cal L}_{5}&=G_5(Y)G_{\m\n}\na^\m A^\n - \ove{6}G'_{5}(Y)
	\big[(\na \cdot A)^{3} -3 (\na \cdot A) \na_{\r} A_{\s}\na^{\s} A^{\r}\\
	&\qquad+2\na_{\r} A_{\s}\na^{\g} A^{\r} \na^{\s} A_{\g}\big]
	- \tilde{G}_5(Y)\Ft^{\a\m}\Ft^\b{}_\m\na_\a A_\m,\\
	{\cal L}_{6}&=G_6(Y)L^{\m\n\a\b}\na_\m A_\n\na_\a A_\b
	+\frac{G'_{6}(Y)}{2}\Ft^{\a\b}\Ft^{\m\n}\na_\a A_\m\na_\b A_\n,
	\end{split}
	\ee
	where $Y\equiv -A_{\m}A^{\m}/2,$ $\na \cdot A\equiv\na_{\m} A^{\m},$
	$\Ft_{\m\n}$ is the Hodge dual of $F_{\m\n},$
	and
	\be
	L^{\m\n\a\b} = \ove{4}\e^{\m\n\r\s}\e^{\a\b\g\d}R_{\r\s\g\d}.
	\ee
	
	Extensions can also be made. In \cite{Heisenberg:2016eld},
	other terms can be added. This results in theories
	called ``beyond generalised Proca theories''.
	In \cite{Kimura:2016rzw}, new classes of theories
	are found and classified by the full analysis of
	vector-tensor theories with up to quadratic order in the first
	derivatives of the vector field.
	The construction made use of ADM decomposition
	to eliminate an unwanted mode.
	In \cite{ErrastiDiez:2019ttn}, \cite{ErrastiDiez:2019trb},
	a system of multiple Maxwell fields
	interacting with multiple Proca fields is constructed.
	This construction is in flat spacetime.
	
	For definiteness,
	let us keep calling the vector sector of the above theories,
	as well as their possible other extensions as generalised Proca theories.
	The construction of the most general generalised Proca theories
	remains an open question.
	In principle, a possible way to do this is
	by following the idea of the original construction of 
	generalised Proca theories,
	that is by starting from demanding that Hessian is
	degenerate. As pointed out recently in \cite{Jimenez:2019hpl},
	in principle, there remains further interaction terms to be found.
	
	We would like to work towards this ultimate goal.
	As an initial step, we take as a hint the usefulness of
	the condition for degenerate Hessian determinant
	in determining generalised Proca theories
	with three propagating degrees of freedom.
	This condition has been used in literature
	with great success since the pioneering works \cite{Tasinato:2014eka}, \cite{Heisenberg:2014rta} on generalised Proca theories.
	So we conjecture that the condition for degenerate Hessian determinant
	almost guarantees that the vector sector has three propagating degrees of freedom.
	This means that we expect that some conditions coming from
	constrained analysis (to get three propagating degrees of freedom)
	are greatly simplified by the condition for degenerate Hessian determinant.
	
	In this paper, we will investigate generalised Proca theories
	which satisfy
	\be\label{Hzeromu}
	\frac{\pa^2\cL}{\pa\dot{A}_0\pa\dot{A}_\m} = 0,\qquad
	\det\lrbrk{\frac{\pa^2\cL}{\pa\dot{A}_i\pa\dot{A}_j}} \neq 0.
	\ee	
	This condition is stronger than the condition that
	Hessian determinant is degenerate.
	Nevertheless, large classes of generalised Proca theories,
	for example the Lagrangian \eq{LgenProca-curved},
	satisfies these conditions.
	The generalised Proca theories that we investigate in this paper
	can be coupled to background metric
	as well as to any other background fields.
	Furthermore, the vector field as well as all the background fields
	should all be diffeomorphic.
	We will argue in this paper that
	diffeomorphism invariance and the condition \eq{Hzeromu}
	almost guarantee that the vector sector has three propagating degrees of freedom.

	This paper is organised as follows.
	In section \ref{sec:method-constraint},
	we give a quick review of the Dirac method
	and the Faddeev-Jackiw method for constrained analysis.
	Then we demonstrate the use of the Faddeev-Jackiw method
	on the Proca theory.
	In section \ref{sec:diffeo-cond}, we derive useful conditions
	from diffeomorphism invariance requirements
	on generalised Proca theories coupled to any background field
	and subject to the condition \eq{Hzeromu}.
	In section \ref{sec:cond-constraint},
	we consider Faddeev-Jackiw analysis of these theories
	and derive the conditions that they should satisfy
	in order for the vector sector to have three propagating degrees of freedom.
	Some these conditions are trivialised by the equations obtained in section \ref{sec:diffeo-cond}.
	As part of cross-checks, we consider explicit examples in section \ref{sec:example}
	and discuss how they satisfy conditions obtained in section \ref{sec:cond-constraint}.
	In section \ref{sec:FJB}, we comment on the implication
	that diffeomorphism invariance has on the Faddeev-Jackiw bracket.
	We conclude this paper and discuss possible future works in section \ref{sec:conclusion}.

	\setcounter{equation}0
	\section{Methods for analysing constrained systems}\label{sec:method-constraint}
	Let us give a quick review of two popular methods
	to analyse constrained systems.
	These methods are the Dirac method and the Faddeev-Jackiw method.
	We give emphases on the latter
	and demonstrate its use on Proca theory.

	\subsection{The Dirac method}\label{subsec:Dirac}	
	In the context of classical field theory,
	a constrained system is the system in which the number of dynamical variables
	is smaller than the number of generalised coordinates.
	For a generalised coordinate to be a dynamical variable,
	its equation of motion should be of second order in time derivative.
	This corresponds to arbitrariness of the initial values of
	the generalised coordinate and the corresponding generalised velocity.
	If, however, the equation of motion of a generalised coordinate
	is at most of first order in time derivative,
	then this equation restricts the initial values
	of the generalised coordinate and velocity.
	This means that the arbitrariness is lost.
	In this case, the generalised coordinate is not a dynamical variable.
	
	The criteria described above can be formalised,
	which gives rise to a simple condition to determine whether a system is constrained.
	In particular, consider a system with the Lagrangian density of the form
	\be\label{L den form_D}
	\cL(\f_a(x),\pa_\m\f_a(x)),
	\ee
	where $a=1,2,\cdots,N.$
	If the determinant of the Hessian
	\be
	\frac{\pa^2\cL}{\pa\dot\f_a\pa\dot\f_b}
	\ee
	is zero, then the system is a constrained system.
	
	In order to analyse constrained systems, the
	Dirac method \cite{Dirac:1950pj}, \cite{dirac2001lectures}, \cite{Anderson:1951ta}
	is a well-known method.
	Starting from the Lagrangian $\cL(\f_a(x),\pa_\m\f_a(x)),$
	one defines conjugate momenta as
	\be\label{conju momem form_D}	
	\p^a = \frac{\pa{\cal L}}{\pa{\dot\f}_{a}}.
	\ee
	If the system is not constrained, eq.\eq{conju momem form_D} can be inverted
	to uniquely express $\dot\f_a$ in terms of $\p^a.$
	But if the system is constrained, this is no longer the case.
	One may then proceed to extract from eq.\eq{conju momem form_D}
	constrained equations, which are equations containing $\f_a$ and $\p^a$
	but without $\dot\f_a.$
	Suppose there are $k$ constrained equations
	of the form $\F^{\hat{m}} = 0,$ for $\hat{m}=1,2,\cdots,k.$
	The quantity $\F^{\hat{m}}$ are called constraints.
	In particular, since they are the initial set of constraints being generated,
	they are called ``primary constraints''.
	
	Next, one requires that the primary constraints should remain constraints
	even after the time has evolved.
	This gives the criteria that the time derivative of constraints should remain on constrained surface in phase space.
	To compute the time derivative, one needs Hamiltonian.
	One obtains the Hamiltonian density by using Legendre transformation giving
	\be\label{H den form_D}	
	{\cal H}=\p^{a}{\dot\f}_{a}-{\cal L} - \g_{\hat{m}}\F^{\hat{m}},
	\ee
	where $\g_{\hat{m}}$ are Lagrange multipliers.
	The Poisson bracket of a phase space variable and the Hamiltonian
	is the time derivative of that variable.
	Suppose that one follows the criteria and obtain further constraints.
	These constraints are called ``secondary constraints''.
	Then the criteria can be applied to the secondary constraints,
	and in case it generate further constraints, these must satisfy the criteria as well.
	The process should be repeated until there are no further constraints generated.
	
	Next, one needs to reclassify all of the constraints of the system.
	If all the Poisson brackets of a constraint with other constraints vanish
	on constrained surfaces,
	then that constraint (sometimes, it is necessary to redefine the constraints
	by writing them as linear combinations of all the constraints) is called a ``first-class constraint''.
	The constraints which are not first-class constraints
	are called second-class constraints.
	Immediate and important usage from this classification
	is to obtain the number of degrees of freedom from the formula
	\be\label{dofformula}
	\textrm{number of d.o.f.} = \frac{n_{PS} - 2n_1 - n_2}{2},
	\ee
	where $n_{PS}$ is the number of phase space variables,
	$n_1$ is the number of first-class constraints,
	and $n_2$ is the number of second-class constraints.
	The counting of degrees of freedom has been useful for example to check whether
	a proposed theory is free of ghost degrees of freedom.
	
	Each class of constraints also have their important roles.
	Let us briefly state them.
	First-class constraints generate gauge transformation.
	As for second-class constraints, they are used
	in order to form Dirac bracket, which is the constrained system
	counterpart of the unconstrained system's Poisson bracket.
	As part of the canonical quantisation of constrained system,
	the Dirac bracket is promoted to commutator.
	
	\subsection{The Faddeev-Jackiw method}\label{subsec:FJ}	
	Another method for analysing the constrained system
	is called the Faddeev-Jackiw method \cite{Faddeev:1988qp}, \cite{Jackiw:1993in}, \cite{BarcelosNeto:1991kw}, \cite{BarcelosNeto:1991ty}.
	This method is relatively simpler than Dirac method,
	for example, one does not need to classify the constraints.	
	Let us give a brief review as follows.
	
	Let us first consider the Lagrangian density in the form of
	of eq.\eq{L den form_D}. Then follow the same discussions
	as in Dirac method until obtaining eq.\eq{H den form_D}.
	However, let us redefine $\g_{\hat{m}}$ as $\dot{\g}_{\hat{m}}.$
	So eq.\eq{H den form_D} becomes
	\be\label{H den form_FJ}	
	{\cal H}=\p^{a}{\dot\f}_{a}-{\cal L} - \dot{\g}_{\hat{m}}\F^{\hat{m}}.
	\ee
	There is no generality lost
	in the redefinition of the Lagrange multiplier.
	Furthermore, it is part of the standard Faddeev-Jackiw algorithm.
	Next, one defines
	\be
	\cL_{FOF}={\p^{a}}{\dot{\f}}_{a}-\cH,
	\label{L_den_FOF_form}
	\ee	
	which is called the first-order form of the Lagrangian.
	The reason for this name is that, by construction, $\cL_{\textrm{FOF}}$ contains terms
	at most of first order derivative in time.
	Then one collects the phase space variables and the constraints
	into the symplectic variables $\xi^I=(\f_a,\p^a,\g_{\hat{m}}),$ for $I=1,2,\cdots,2N+k.$
	Then one defines
	\bea
	{\cal A}_{{\xi}^{I}}=\frac{\pa {\cal L}_{FOF}}{\pa \dot\xi^{I}}.
	\label{cano momen_form}
	\eea	
	Note that we have used the symplectic variables $\xi^I$ as labels
	instead of just simply the indices $I.$
	Next, by using eq.\eq{cano momen_form}, $\cL_{\textrm{FOF}}$ can be written in the form
	\bea
	{\cal L}_{FOF}={\cal A}_{{\xi}^{I}}{\dot {\xi}^{I}}+{\cal L}_{v}.
	\label{L_FOF_den_form_new}
	\eea
	Note that $\cA_{\xi^I}$ and $\cL_v$ only depend on $\xi^I$ but not $\dot{\xi}^I.$
	
	In order to proceed, it is convenient to make use of differential form language.
	Let us denote the coordinate basis for vector and $1$-form in the space of $\xi^I$
	as
	\be
	\frac{\d}{\d\xi^I(t,\vec{x})},\qquad\textrm{and}\qquad \d\xi^I(t,\vec{x}),
	\ee
	respectively. Let us define the one-form corresponding to $\cA_{\xi^I}$ as
	\be
	\cA(t) \equiv \int d^3\vec{x}\ \cA_{\xi^I}(t,\vec{x})\d\xi^I(t,\vec{x}).
	\ee
	The quantity $\cA(t)$ is called the canonical $1$-form.
	Applying the exterior derivative
	\be
	\d \equiv
	\int d^3\vec{x}\ \d\xi^I(t,\vec{x})\frac{\d}{\d\xi^I(t,\vec{x})}
	\ee
	on $\cA(t)$ gives
	\be
	\begin{split}
	\cF(t)
	&\equiv\d\cA(t)\\
	&=\hlf\int d^3\vec{x}\int d^3\vec{y}\
	\lrbrk{\frac{\d\cA_{\xi^J}(t,\vec{y})}{\d\xi^I(t,\vec{x})} - \frac{\d\cA_{\xi^I}(t,\vec{x})}{\d\xi^J(t,\vec{y})}}
	\d\xi^I(t,\vec{x})\wedge\d\xi^J(t,\vec{y})\\
	&\equiv\hlf\int d^3\vec{x}\int d^3\vec{y}\
	\cF_{\xi^I\xi^J}(t,\vec{x},\vec{y})
	\d\xi^I(t,\vec{x})\wedge\d\xi^J(t,\vec{y}).
	\end{split}
	\ee
	The quantity $\cF(t)$ is called the symplectic $2$-form.
	It is an important quantity which is used to determine
	if the system has additional constraints.
	For this, let us consider an equation
	\be\label{zero mode}
	i_{z(t)}\cF(t) = 0,
	\ee
	where
	\be
	z(t) = \int d^3\vec{x}\ z^{\xi^I}(t,\vec{x})\frac{\d}{\d\xi^I(t,\vec{x})}.
	\ee
	The non-trivial solution $z^{\xi^I}(t,\vec{x})$
	to eq.\eq{zero mode} is just the zero mode of $\cF_{\xi^I\xi^J}(t,\vec{x},\vec{y}).$
	If there is no zero mode, then the system has no further constraint.
	However, if zero mode exists, there might be new constraints.
	The new constraints are generated from
	\be\label{constraint-gen}
	\Omega(t)=i_{z(t)}\d\int d^3\vec{x}\ {\cal L}_{v}(t,\vec{x}),
	\ee
	where $z(t)$ is the non-trivial solution to eq.\eq{zero mode}.
	It might turn out that
	some zero modes do not lead to a new constraint.
	This case can occur if after substituting these zero modes
	into eq.\eq{constraint-gen},
	one obtains some trivial conditions (for example $\Omega = 0$),
	or constraints dependent on the ones already discovered.
	
	In case there are new constraints generated from the above steps,
	one needs to repeat the above steps by first modifying the first-order Lagrangian.
	The iterations should be continued until there is no further constraint
	obtained. For definiteness, let us call the steps we just discussed as the first iteration.
	Next, let us discuss how the next iterations should be carried out.
	
	Suppose there are $k'$ new constraints.
	Let us denote all constraints so far as $\F_{\hat{m}},$
	where the index $\hat{m}$ is redefined to take values $\hat{m} = 1,2,\cdots, k+k'.$
	The Hamiltonian is then redefined accordingly so that it still takes the form of eq.\eq{H den form_FJ},
	but with the range of $\hat{m}$ changed to $1,2,\cdots, k+k'.$
	The new first-order Lagrangian can be obtained from the new Hamiltonian
	by using eq.\eq{L_den_FOF_form}.
	Consequently, eq.\eq{cano momen_form} and eq.\eq{L_FOF_den_form_new} still apply,
	but with symplectic variables redenoted as $\xi^I\equiv(\f_a,\p^a,\g_{\hat{m}}),$
	for $I = 1,2,\cdots,2N+k+k'.$ One may then follow the steps in the previous paragraph
	to obtain further constraints. If there is a new one, repeat the steps again and again
	until there is no further constraint generated.
	
	When the iterations end, the matrix inverse of the symplectic $2-$form
	gives Faddeev-Jackiw bracket $\FJB{.}{.}.$
	That is \cite{Jackiw:1993in}
	\be
	\FJB{\xi^I(t,\vec{x})}{\xi^J(t,\vec{y})} = (\cF^{-1})^{\xi^I\xi^J}(t,\vec{x},\vec{y}),
	\ee
	where $(\cF^{-1})^{\xi^I\xi^J}(t,\vec{x},\vec{y})$ is the matrix inverse of $\cF_{\xi^I\xi^J}(t,\vec{x},\vec{y})$
	in the sense that
	\be
	\int d^3\vec{y}\cF_{\xi^I\xi^J}(t,\vec{x},\vec{y})(\cF^{-1})^{\xi^J\xi^K}(t,\vec{y},\vec{z})
	=\d_{I}^K\d^{(3)}(\vec{x}-\vec{z}).
	\ee
	It has been argued in literature \cite{Jackiw:1993in},
	that in various theories,
	Faddeev-Jackiw bracket is equivalent to Dirac bracket.
	So after canonical quantisation, the Faddeev-Jackiw bracket
	could be promoted to commutator.
	
	\subsection{Faddeev-Jackiw analysis of Proca theory}
	Let us review the analysis of Proca theory using the Faddeev-Jackiw method.
	This analysis will form a basis for later generalisations
	performed in this paper.
	
	The Proca action in a four-dimensional flat Minkowski spacetime with signature $(-,+,+,+)$ is given by
		\be\label{S_Proca}
		S_{Proca}=\int d^{4}x\lrbrk{-\frac{1}{4}F_{\mu\nu}F^{\mu\nu}-\frac{1}{2}m^{2}A_{\m}A^{\m}},
		\ee
	where $F_{\m\n}=\pa_\m A_\n - \pa_\n A_\m$ with $\m,\n = 0,1,2,3.$
	The conjugate momenta to $A_\m$ can easily be worked out,
	and are given by
		\be\label{conju momen_Proca}
		\p^{\m}=F^{\m0}.
		\ee
	From these equations, one obtains the primary constraint
	\be\label{con EQ_Proca}
	\Omega_1\equiv\p^{0}=0.	
	\ee 
	The Hamiltonian density is given by
	\be\label{H den_FOF_Proca}
	{\cal H}(A^{\r},{\p^{\r}})
	=\frac{1}{2}\p^i\p^{i}+{\p^{i}}(\pa_{i}A_{0})
	+\frac{1}{2}m^{2}A_\m A^\m+\frac{1}{4}F_{ij}F^{ij}-\dot{\g}_1{\p^{0}},
	\ee
	where $\dot{\g}_1$ is a Lagrange multiplier,
	and $A^{2}\equiv A_{\m}A^{\m}.$
	This gives
	\be\label{L_FOF_Proca}
	\cL_{FOF}={\p^{\r}}{\dot{A_{\r}}}-\frac{1}{2}{\p^{i}\p^i}-{\p^{i}}(\pa_{i}A_{0})-\frac{1}{2}m^{2}A^{2}-\frac{1}{4}F_{ij}F^{ij}+\dot{\g}_1{\p^{0}}.
	\ee
	After putting eq.\eq{L_FOF_Proca} in the form of
	eq.\eq{L_FOF_den_form_new}, one sees that ${\cal L}_{v}$ is given by
	\be\label{Lv Proca}
	{\cal L}_{v}=-\frac{1}{2}{\p^{i}\p^{i}}-{\p^{i}}(\pa_{i}A_{0})-\frac{1}{2}m^{2}A_\m A^{\m}-\frac{1}{4}F_{ij}F^{ij}.
	\ee
	Let us define the symplectic variables as $\xi^{I}=(A_{0}, A_{i}, \p^{0}, \p^{i}, \g_1),$
	and compute the canonical momenta using eq.\eq{cano momen_form}.
	The corresponding canonical $1-$form is given by
	\be\label{cano 1-form_Proca}
	{\cal A}=\int d^{3}\vec{x} \lrbrk{\p^{0}\d A_{0}+{\p^{i}}\d A_{i}+{\p^{0}}\d \g_1}.
	\ee
	The symplectic $2-$form is then	
	\be\label{sym 2 form_Proca_1} 
	{\cal F}=\int d^{3}\vec{x} \lrbrk{\d\p^{0}\w\d A_{0}+\d{\p^{i}}\w\d A_{i}+\d\p^{0}\w\d \g_1 }.
	\ee
	This gives
	\be
	\label{zero mode_Proca_1} 
	i_z{\cal F}
	=\int d^{3}\vec{x} \Big(z^{\p^{0}}\d A_{0}-z^{ A_{0}}\d\p^{0}+ z^{A_{i}}\d{\p^{i}}-z^{\p^{i}}\d A_{i}
	+z^{\p^{0}}\d \g_1-z^{\g_1}\d\p^{0} \Big).
	\ee
	The solution to $i_z\cF = 0$ is then
	\be\label{zProca-1}
	z = \int d^3\vec{x}\ z^{A_0}\lrbrk{\frac{\d}{\d A_0} - \frac{\d}{\d \g_1}}.
	\ee
	By using eq.\eq{constraint-gen},
	\eq{Lv Proca}, and \eq{zProca-1},
	one sees that there is an additional constraint given by
	\be\label{constraint_Proca_gamma1} 
	\Omega_{2}=\pa_{i}\p^{i}+m^{2}A_{0}.
	\ee
	
	Due to the presence of the new constraint \eq{constraint_Proca_gamma1},
	one needs to consider the second iteration by starting from
	the following first-order form of Lagrangian
	\be\label{L_FOF_Proca_new}
	{\cal L}_{FOF} ={\p^{\r}}{\dot{A_{\r}}}-\frac{1}{2}{\p^{i}\p^{i}}
	-{\p^{i}}(\pa_{i}A_{0})-\frac{1}{2}m^{2}A^{\m}A_\m
	-\frac{1}{4}F_{ij}F^{ij}+\dot{\g}_1{\p^{0}}+\dot{\g}_{2}(\pa_{i}\p^{i}+m^{2}A_{0}).
	\ee
	Furthermore, the symplectic variables are now
	$\xi^{I}=(A_{0}, \p^{0}, A_{i}, \p^{i}, \g_1, \g_{2}) $.
	The canonical $1-$form and the symplectic $2-$form are then given by
	\be\label{cano 1-form_Proca_2new}
	{\cal A}
	=\int d^{3}\vec{x} \lrbrk{\p^{0}\d A_{0}+\p^{i}\d A_{i}+\p^{0}\d \g_1
	 +(\pa_{i}\p^{i}+m^{2}A_{0})\d \g_{2} },
	\ee
	and
	\be\label{sym 2 form_Proca_2new} 
	{\cal F}=\int d^{3}\vec{x} \lrbrk{\d\p^{0}\w\d A_{0}+\d\p^{i}\w\d A_{i}+\d\p^{0}\w\d \g_1
	+(\pa_{i}\d\p^{i}+m^{2}{\d A_{0}})\w\d \g_{2} }.
	\ee
	Applying interior derivative on eq.\eq{sym 2 form_Proca_2new} gives
	\be\label{sym 2 form_Proca_2super new} 
	\begin{split}
	i_z\cF&=\int d^{3}\vec{x} \Big(
	(z^{\p^{0}}-m^{2}z^{\g_{2}})\d A_{0}+z^{\p^{i}} \d A_{i}
	-(z^{A_{0}}+z^{\g_1})\d\p^{0}+(\pa_{i}z^{\g_{2}}-z^{A_{i}})\d{\p^{i}}\\
	&\qquad\qquad\qquad+z^{\p^{0}}\d\g_1+(\pa_{i}z^{\p_{i}}+m^{2}z^{A_{0}})\d\g_{2} \Big).
	\end{split}
	\ee
	The equation $i_z\cF = 0$ can be solved step-by-step as follows.
	One starts from considering coefficients of $\d A_i$ and $\d\g_1.$
	This gives $z^{\p^\m} = 0.$
	Next, one considers coefficients of $\d A_0$ and $\d\g_2.$
	This gives $z^{\g_2} = 0 = z^{A_0}.$
	Finally, one considers coefficients of $\d \p^\m.$
	This gives $z^{\g_1} = 0 = z^{A_i}.$
	Therefore, the only solution is $z=0,$
	and hence there is no further constraint.
	
	With all the constraints at hand,
	it is now possible to count the number of degrees of freedom.
	This is by first noting that the Poisson's bracket
	between the two constraints is
	\be\label{PB-Proca}
	\PB{\p^0(\vec{x})}{(\pa_i\p^i + m^2 A_0)(\vec{y})} = -m^2\d^{(3)}(\vec{x}-\vec{y}).
	\ee
	Note that we have omitted writing the dependence on $t.$
	So for example $\p^0(\vec{x})$ stands for $\p^0(t,\vec{x}).$
	From eq.\eq{PB-Proca}, we see that
	both constraints are of second-class.
	The number of degrees of freedom can be obtained from eq.\eq{dofformula}.
	In this case, $n_{PS} = 8, n_1 = 0,$ and $n_2 = 2.$
	Therefore the Proca theory has three degrees of freedom.
	
	The analysis given above is for the case $m\neq 0.$
	The analysis for the care $m=0$ has to be given separately.
	For this, we may easily follow the steps from eq.\eq{S_Proca}
	to eq.\eq{sym 2 form_Proca_2super new} by simply setting $m=0.$
	So at this stage, there are in total
	two constraints: $\p^0 \approx 0$ and $\pa_i\p^i\approx 0.$
	The zero modes to eq.\eq{sym 2 form_Proca_2super new}
	with $m=0$ are given by
	\be\label{zProca-Maxwell}
	z = \int d^3\vec{x}\ \lrbrk{z^{A_0}\lrbrk{\frac{\d}{\d A_0} - \frac{\d}{\d \g_1}}
	+z^{\g_2}\frac{\d}{\d\g_2} + \pa_i z^{\g_2}\frac{\d}{\d A_i}}.
	\ee	
	Although there are two zero modes,
	it can be checked that they do not lead to a new constraint.
	So the process has to stop.
	The Poisson's bracket between the two constraints
	can then be computed and found that it vanishes:
	\be\label{PB-Maxwell}
	\PB{\p^0(\vec{x})}{(\pa_i\p^i)(\vec{y})} = 0.
	\ee
	So both of the constraints are of first-class,
	and hence the theory has two degrees of freedom.
	
	In fact, it can already be seen
	from the Faddeev-Jackiw iterative process, without computing Poisson's bracket,
	that there exists first-class constraints for Maxwell theory
	but there is no first-class constraint in Proca theory.
	The criteria is provided by \cite{Rodrigues:2018ioe}, which is summarised as follows.
	If all of the zero modes in the Faddeev-Jackiw process
	give rise to independent constraints,
	then there is no first-class constraint.
	However, if there are zero modes which do not give rise
	to new constraints, then there are first-class constraints.
	In fact, \cite{Rodrigues:2018ioe} also provides the way
	of counting number of degrees of freedom
	from the number of zero modes and of constraints.
	However, we will not discuss this way of counting in this paper.
	
	Next, let us obtain Faddeev-Jackiw bracket.
	We first read off the matrix $\cF_{\xi^I\xi^J}(\vec{x},\vec{y})$
	from eq.\eq{sym 2 form_Proca_2new}.
	We have
	\be\label{cF-Proca-matrix}
	\cF_{\xi^I\xi^J}(t,\vec{x},\vec{y})
	=
	\begin{pmatrix}
	0 & -\d^\m{}_\s & 0 & m^2\d^\m{}_0\\
	\d_\n{}^\r & 0 & \d_\n{}^0 & -\d_\n{}^i\pa_{x^i}\\
	0 & -\d^0{}_\s & 0 & 0\\
	-m^2\d_0{}^\r & -\d^i{}_\s\pa_{x^i} & 0 & 0
	\end{pmatrix}
	\d^{(3)}(\vec{x}-\vec{y}),
	\ee
	where we have used the labels
	$\xi^I(t,\vec{x}) = (A_\m(t,\vec{x}), \p^\n(t,\vec{x}), \g_1(t,\vec{x}), \g_2(t,\vec{x})),$
	and $\xi^J(t,\vec{y}) = (A_\r(t,\vec{y}), \p^\s(t,\vec{y}), \g_1(t,\vec{y}), \g_2(t,\vec{y})).$
	Inverting eq.\eq{cF-Proca-matrix} gives
	\be
	(\cF^{-1})^{\xi^J\xi^K}(t,\vec{x},\vec{y})
	=
	\begin{pmatrix}
	\frac{2}{m^2}\d^0_{(\r}\d^i_{\k)}\pa_{x^i} & \d_{\r}^i\d^\l_i & -\d_{\r}^i\frac{\pa_{y^i}}{m^2} & -\frac{\d_{\r}^0}{m^2}\\
	-\d^\s_i\d_{\k}^i & 0 & -\d^\s{}_0 & 0\\
	-\d_{\k}^i\frac{\pa_{y^i}}{m^2} & \d_0{}^\l & 0 & \ove{m^2}\\
	\frac{\d^0{}_\k}{m^2} & 0 & -\ove{m^2} & 0
	\end{pmatrix}
	\d^{(3)}(\vec{y}-\vec{z}).
	\ee	
	So the Faddeev-Jackiw brackets between the canonical variables are
	\bea
	\FJB{A_\m(t,\vec{x})}{A_\r(t,\vec{y})} &=& \frac{2}{m^2}\d^0_{(\m}\d^i_{\r)}\pa_{x^i}\d^{(3)}(\vec{x}-\vec{y}),\\
	\FJB{A_\m(t,\vec{x})}{\p^\s(t,\vec{y})} &=& \d_{\m}^i\d^\s_i\d^{(3)}(\vec{x}-\vec{y}),\\
	\FJB{\p^\r(t,\vec{x})}{\p^\s(t,\vec{y})} &=& 0,
	\eea
	which, after taking into account the different convention for metric signature,
	is in exact agreement with Dirac's bracket given in \cite{Vytheeswaran:1997jr}, \cite{Kim:1996gk}.
	
	\setcounter{equation}0	
	\section{Conditions from diffeomorphism invariance}\label{sec:diffeo-cond}
	A generalised Proca Lagrangian which satisfies the requirement \eq{Hzeromu}
	should take the form
	\be\label{LGP-UV}
	\begin{split}
	\cL
	&= T(A_\n,\pa_k A_\n, \dot A_k,g_{\r\s},\pa_{\k}g_{\r\s},\pa_{\k\l}g_{\r\s},\cdots,K)\\
	&\qquad+ U(A_\n,\pa_i A_\n,g_{\r\s},\pa_{\k}g_{\r\s},\pa_{\k\l}g_{\r\s},\cdots,K) \dot{A}_0.
	\end{split}
	\ee
	Here, only $T$ is allowed to depend on $\dot{A}_k,$
	but $U$ should not be.
	Note that the action \eq{LGP-UV} describes the dynamics of a vector field coupled to
	the background metric $g_{\m\n}$
	and their derivatives.
	Furthermore we also include for completeness,
	the coupling to other external fields which are,
	along with their possible derivatives of any order,
	collectively called $K.$

	The Lagrangian \eq{LGP-UV} is free of Ostrogradsky instability \cite{Ostrogradsky:1850fid}.
	This is because the Lagrangian depends only up to the first
	order in time derivative of the field $A_\m.$
	On the other hand, the appearance of time derivative of the metric
	and of other external fields in eq.\eq{LGP-UV}
	need not concern us because the dynamics of these fields
	are not determined by the theory \eq{LGP-UV}.
	Of course, care must be taken when this Lagrangian is included
	into the full Lagrangian, in which every field is dynamical.
	We leave this as a future work. See also section \ref{sec:conclusion}.
	
	We also require that the theory \eq{LGP-UV}
	is invariant under diffeomorphism.
	This requirement will impose conditions on
	the form of $T$ and $U.$
	In practice, one would normally wish to
	start from a Lagrangian which is already diffeomorphism invariance,
	and then put it in the form \eq{LGP-UV}
	and proceed with constrained analysis.
	So it seems conditions coming from
	diffeomorphism invariance requirement do not need to be stated at all.
	It turns out, however, that
	some conditions coming from
	diffeomorphism invariance requirement
	will help simplifying the constrained analysis.
	So we will derive these useful
	conditions before working on
	constrained analysis.

	In this section, we are going to consider diffeomorphism
	transformation on the Lagrangian \eq{LGP-UV}
	and require that the theory is diffeomorphism invariant.
	The conditions to be used do not depend explicitly on $g_{\m\n}$
	and other external fields.
	However, to demonstrate the calculation,
	we let $K$ in eq.\eq{LGP-UV}
	to be a collection $(\pa_{\s_1\cdots\s_p}k^{\m_1\cdots\m_m}{}_{\n_1\cdots\m_n}: m,n,p=0,1,\cdots),$
	where $k^{\m_1\cdots\m_m}{}_{\n_1\cdots\n_n}$ is an external tensor field
	of rank $(m,n).$ Of course, the results to be found in this section
	also hold when $K$ also include external fermion fields.
	
	Consider a diffeomorphism transformation $x^\m\to x^\m-\e^\m(x),$
	where $\e^\m(x)$ are arbitrary functions of spacetime.
	Under this transformation, the fields transform as Lie derivative:
	\be
	\d_\e A_\m = \cL_\e A_\m = \e^\r\pa_\r A_\m + A_\r\pa_\m \e^\r,
	\ee
	\be
	\d_\e g_{\m\n}
	= \cL_\e g_{\m\n} = \e^\r\pa_\r g_{\m\n} + 2g_{\r(\n}\pa_{\m)} \e^\r,
	\ee
	\be
	\begin{split}
	\d_\e k^{\m_1\cdots\m_m}{}_{\n_1\cdots\n_n}
	&= \cL_\e k^{\m_1\cdots\m_m}{}_{\n_1\cdots\n_n}\\
	&= \e^\r\pa_\r k^{\m_1\cdots\m_m}{}_{\n_1\cdots\n_n}\\
	&\qquad-k^{\r\m_2\cdots\m_m}{}_{\n_1\cdots\n_n}\pa_{\r}\e^{\m_1}
	-\cdots
	-k^{\m_1\cdots\m_{m-1}\r}{}_{\n_1\cdots\n_n}\pa_{\r}\e^{\m_m}\\
	&\qquad+k^{\m_1\cdots\m_m}{}_{\r\n_2\cdots\n_l}\pa_{\n_1}\e^\r
	+\cdots
	+k^{\m_1\cdots\m_m}{}_{\n_1\cdots\n_{n-1}\r}\pa_{\n_n}\e^\r.
	\end{split}
	\ee
	The field variation $\d_\e$ satisfies Leibniz rules.
	Furthermore, it also commutes with partial derivatives.
	So for example,
	\be
	\d_\e \dot{A}_\m
	=\pa_0(\e^\r\pa_\r A_\m + A_\r\pa_\m \e^\r)
	=\dot{\e}^\r\pa_\r A_\m + \e^\r\pa_\r \dot{A}_\m + \dot{A}_\r\pa_\m \e^\r
	+A_\r\pa_\m \dot{\e}^\r,
	\ee
	\be
	\begin{split}
	\d_\e T
	&=\frac{\pa T}{\pa A_\n}\d_\e A_\n + \frac{\pa T}{\pa\pa_k A_\n}\pa_k \d_\e A_\n
	+\frac{\pa T}{\pa\dot{A}_k}\pa_0 \d_\e A_k\\
	&\qquad+ \frac{\pa T}{\pa g_{\m\n}}\d_\e g_{\m\n}
	+ \frac{\pa T}{\pa\pa_\s g_{\m\n}}\pa_\s \d_\e g_{\m\n} + \cdots\\
	&\qquad+ \frac{\pa T}{\pa k}\d_\e k + \frac{\pa T}{\pa \pa_\m k}\pa_\m\d_\e k
	+\cdots\\
	&\qquad+\frac{\pa T}{\pa \pa_{\s_1\cdots\s_p}k^{\m_1\cdots\m_m}{}_{\n_1\cdots\m_n}}
	\pa_{\s_1\cdots\s_p}\d_\e k^{\m_1\cdots\m_m}{}_{\n_1\cdots\m_n}+\cdots.
	\end{split}
	\ee
	
	For the theory to be diffeomorphism invariant,
	the Lagrangian should transform as
	\be\label{dcLcond}
	\d_\e\cL = \e^\m\pa_\m\cL + \cL\pa_\m\e^\m.
	\ee
	This is the requirement that the Lagrangian \eq{LGP-UV} has to satisfy.
	After making $\d_\e$ variation on the Lagrangian \eq{LGP-UV},
	one can see that the coefficient of $\e^\m$
	is already $\pa_\m\cL.$
	Furthermore, it can be seen that
	$\d_\e\cL - \pa_\m(\e^\m\cL)$
	can be given as a quadratic polynomial in $\dot A_0$
	(it needs not be a polynomial in $\dot A_i$).
	So coefficient of each order in $\dot A_0$ should vanish.
	This gives three equations.
	Each of them can furthermore be expressed
	as linear combinations of expressions of the form
	$\pa_{\s_1\cdots\s_p}\e^\m.$
	So coefficient of each of these expressions should vanish.

	Let us consider the coefficient $\dot{A}_0\dot{A}_0.$
	It turns out there is only one term contributes to this coefficient.
	That is
	\be
	\frac{\pa U}{\pa\pa_i A_0}\pa_i\e^0\dot{A}_0\dot{A}_0\subset\d_\e\cL - \pa_\m(\e^\m\cL).
	\ee
	This implies that
	\be\label{condU0}
	\frac{\pa U}{\pa\pa_i A_0} = 0.
	\ee
	Let us now turn to the coefficient of $\dot A_0.$
	We have
	\be\label{condpaxidotA0}
	\lrbrk{\frac{\pa T}{\pa\pa_k A_0}\pa_k\e^0+\frac{\pa T}{\pa\dot{A}_k}\pa_k\e^0
	+2U\pa_0\e^0 + U\pa_\m\e^\m
	+\d_\e U\bigg|_{\dot{A}_0 = 0}}\dot{A}_0\subset\d_\e\cL - \e^\m\pa_\m\cL,
	\ee
	where the coefficient of $\dot{A}_0$ on the last term on LHS
	are obtained from $\d_\e U$ by setting $\dot{A}_0 = 0$
	and discarding terms linear in $\e^\m.$
	Conditions can be extracted by letting LHS of eq.\eq{condpaxidotA0}
	to be zero and demanding the coefficients of $\dot{A}_0$
	to vanish. In particular, we are interested in the coefficient of $\pa_k\e^0.$
	So we have
	\be\label{TUfull}
	\begin{split}
	0&=\frac{\pa T}{\pa\pa_k A_0}+\frac{\pa T}{\pa\dot{A}_k}
	+\frac{\pa U}{\pa A_k}A_0 + \frac{\pa U}{\pa\pa_k A_i}\dot A_i
	+\frac{\pa U}{\pa\pa_i A_k}\pa_i A_0\\
	&\qquad+2\frac{\pa U}{\pa g_{\m k}}g_{\m0}
	+\frac{\pa U}{\pa\pa_k g_{\r\s}}\pa_0 g_{\r\s}
	+2\frac{\pa U}{\pa\pa_\m g_{\r k}}\pa_\m g_{\r 0}+\cdots,
	\end{split}
	\ee
	where $\cdots$ are terms involving partial derivatives of $U$
	with respect to higher order derivatives of $g_{\m\n}$
	and to external tensor fields and their derivatives.
	Two useful conditions can readily be extracted.
	One of them is obtained by taking derivative of
	eq.\eq{condpaxidotA0}
	with respect to $\pa_j A_0,$
	and use eq.\eq{condU0}.
	One obtains
	\be\label{TTU}
	\frac{\pa^2 T}{\pa\dot A_i\pa\dot A_j} + \frac{\pa^2 T}{\pa\dot A_i\pa\pa_j A_0}
	+\frac{\pa U}{\pa\pa_j A_i} = 0.
	\ee
	Similarly, the other condition we are interested in
	can be obtained by taking derivative of eq.\eq{condpaxidotA0}
	with respect to $\pa_j A_0,$
	and use eq.\eq{condU0}.
	One obtains
	\be\label{condTV}
	\frac{\pa^2 T}{\pa\pa_i A_0\pa\pa_j A_0}
	+\frac{\pa^2 T}{\pa\dot{A}_i\pa\pa_j A_0}
	+\frac{\pa U}{\pa\pa_j A_i} = 0.
	\ee	
	There is no further condition that will be of use for us.
	So we end the analysis of diffeomorphism invariance here.

	\setcounter{equation}0
	\section{Conditions from Faddeev-Jackiw constrained analysis of generalised Proca theories}\label{sec:cond-constraint}
	\subsection{Deriving the conditions}
	Let us now proceed by
	considering the Faddeev-Jackiw constrained analysis of the Lagrangian \eq{LGP-UV}.
	Let us first compute the
	conjugate momenta. It is found to be of the form
	\be
	\begin{split}
	\p^\m
	&= \frac{\pa\cL}{\pa\dot{A}_\m}\\
	&= \frac{\pa T}{\pa\dot A_k}\d^\m_k + U\d^\m_0.
	\end{split}
	\ee
	The zeroth component gives the constraint
	\be\label{constraint1}
	\Omega_1\equiv \p^0 - U = 0,
	\ee
	which does not depend on time derivative of the field $A_\m.$
	As for the spatial components for conjugate momenta,
	they give
	\be
	\p^k = \frac{\pa T}{\pa\dot A_k}.
	\ee
	The inverse of this equation is of the form
	\be
	\dot A_i = \L_i(A_\n,\pa_k A_\n,\p^k,
	g_{\r\s}, \pa_{\k}g_{\r\s},\pa_{\k\l}g_{\r\s},\cdots,K).
	\ee
	So the first-order form of Lagrangian density is given by
	\be\label{cLFOFGP}
	\cL_{FOF} = \p^0\dot{A}_0 + \p^i\dot{A}_i + \cL_v + \dot{\g}_1\Omega_1,
	\ee
	where $\cL_v$ is given by	
	\be
	\cL_v = -\p^k\L_k + \cT,
	\ee
	where $\cT$ is obtained by replacing $\dot A_i$ in $T$ by $\L_i.$
	The canonical variables are $\xi^I = (A^\m,\p_\n,\g_1).$
	By using eq.\eq{cano momen_form}, one obtains the canonical $1-$form as
	\be
	\cA= \int d^3\vec{x}\lrbrk{ {\p^{0}}\d A_{0}
	+\p^{i}\d A_{i} + \Omega_1\d\g_1}.
	\ee
	The symplectic $2-$form $\cF = \d\cA$ can then be computed,
	and is given by
	\be
	\begin{split}
	\cF
	&= \int d^3\vec{x}
	\bigg(\d\p^0\w\d A_0 + \d\p^i\w\d A_i
	+\frac{\pa\Omega_1}{\pa A_\m}\d A_\m\w\d\g_1\\
	&\qquad\qquad\qquad+\frac{\pa\Omega_1}{\pa\pa_i A_\m}\d\pa_i A_\m\w\d\g_1
	+\d\p^0\w\d\g_1
	\bigg).
	\end{split}
	\ee
	Then an interior product with a vector $z$ is given by
	\be\label{izcF-1}
	\begin{split}
	i_z\cF
	&=\int d^3\vec{x}
	\bigg(\lrbrk{z^{\p^0} - \frac{\pa\Omega_1}{\pa A_0}z^{\g_1} + \pa_i\lrbrk{\frac{\pa\Omega_1}{\pa\pa_i A_0}z^{\g_1}}}\d A_0\\
	&\qquad\qquad\qquad+\lrbrk{z^{\p^i} - \frac{\pa\Omega_1}{\pa A_i}z^{\g_1} + \pa_j\lrbrk{\frac{\pa\Omega_1}{\pa\pa_j A_i}z^{\g_1}}}\d A_i\\
	&\qquad\qquad\qquad+(-z{^{A_0}} - z^{\g_1})\d\p^0 -z^{A_i}\d\p^i\\
	&\qquad\qquad\qquad+\lrbrk{\frac{\pa\Omega_1}{\pa A_\m}z^{A_\m} + \frac{\pa\Omega_1}{\pa\pa_i A_\m}\pa_iz^{A_\m} + z^{\p^0}}\d\g_1
	\bigg),
	\end{split}
	\ee
	where partial derivatives for $\Omega_1$ are taken
	on $\Omega_1$ of the form
	\be
	\Omega_1(A_\m,\pa_i A_\m,\p^\m,g_{\m\n},\pa_\s g_{\m\n},\cdots,K).
	\ee	
	We wish to obtain the solution to $i_z\cF = 0.$
	For this, let us first consider the coefficients of $\d A_0, \d\p^0$ and $\d\g_1.$
	The condition that these coefficients vanish gives
	\be\label{zA0-eqn}
	2\frac{\pa\Omega_1}{\pa\pa_i A_0}\pa_i z^{A_0} + \pa_i\lrbrk{\frac{\pa\Omega_1}{\pa\pa_i A_0}} z^{A_0} = 0.
	\ee
	In the analysis so far in this subsection,
	we still have not used diffeomorphism invariance requirement.
	In particular, let us impose eq.\eq{condU0}.
	This makes eq.\eq{zA0-eqn} identically vanishes,
	and hence it does not give a restriction on $z^{A_0}.$
	Furthermore, the condition $i_z\cF = 0$
	can be consistently solved,
	and the zero mode of $\cF$ is found to be
	\be
	\begin{split}
	z&=\int d^3\vec{x}\bigg(-\frac{\pa\Omega_1}{\pa A_0}z^{A_0}\frac{\d}{\d\p^0}
	-\frac{\pa\Omega_1}{\pa A_i}z^{A_0}\frac{\d}{\d\p^i}\\
	&\qquad\qquad\qquad+\pa_j\lrbrk{\frac{\pa\Omega_1}{\pa\pa_j A_i}z^{A_0}}\frac{\d}{\d\p^i}
	+z^{A_0}\lrbrk{\frac{\d}{\d A_0} - \frac{\d}{\d\g_1}}\bigg).
	\end{split}
	\ee
	Since the zero mode depend only on one arbitrary function $z^{A_0},$
	there is at most one new constraint.
	It turns out that indeed there is a further constraint $\Omega_2,$
	which can be obtained from
	\be
	\int d^3\vec{x}\ \Omega_2z^{A_0} = i_z\int d^3\vec{x}\ \d\cL_{v}.
	\ee
	After a direct calculation, we obtain
	\be\label{constraint2}
	\begin{split}
	\Omega_2
	&=\frac{\pa \cT}{\pa A_0}
	- \pa_i\lrbrk{\frac{\pa \cT}{\pa\pa_i A_0}}
	-\L_i\frac{\pa U}{\pa A_i} - \pa_j\L_i\frac{\pa U}{\pa\pa_j A_i},
	\end{split}
	\ee
	where partial derivatives of $\cT$ with respect to $A_0$ and $\pa_i A_0$
	are taken with fixed $\L_j.$
	It is easy to see that the constraint \eq{constraint2}
	is indeed a new constraint. This is because it is independent from $\p^0,$
	which appears in $\Omega_1.$
	
	So we have seen an interesting result
	that diffeomorphism invariance
	ensures that the theory \eq{LGP-UV}
	has more than one constraint.
	There will be a further result,
	which we will encounter shortly.

	With the introduction of an extra constraint,
	we need to start the second iteration by first
	redefining
	the first-order form of Lagrangian
	from eq.\eq{cLFOFGP} to
	\be\label{cLFOFGP-2}
	\cL_{FOF} = \p^0A_0 + \p^iA_i + \cL_v + \dot{\g}_1\Omega_1
	+ \dot{\g}_2\Omega_2.
	\ee
	Then the canonical $1-$form for the Lagrangian \eq{cLFOFGP-2}
	is given by
	\be
	\cA= \int d^3\vec{x}\lrbrk{ {\p^{0}}\d A_{0}
	+\p^{i}\d A_{i} + \Omega_1\d\g_1 + \Omega_2\d\g_2}.
	\ee
	Then, we obtain
	\be\label{CF-2ndit}
	\cF = \int d^3\vec{x}\lrbrk{\d\p^0\w\d A_0 + \d\p^i\w\d A_i
	+\d\Omega_1\w\d\g_1 + \d\Omega_2\w\d\g_2},
	\ee
	and hence
	\be\label{izcF-2}
	\begin{split}
	i_z\cF
	&=\int d^3\vec{x}
	\Bigg(\bigg(z^{\p^0} - \frac{\pa\Omega_1}{\pa A_0}z^{\g_1}
	- \frac{\pa\Omega_2}{\pa A_0}z^{\g_2}\\
	&\qquad\qquad\qquad\qquad\qquad+ \pa_j\lrbrk{\frac{\pa\Omega_2}{\pa\pa_j A_0}z^{\g_2}}
	- \pa_j\pa_k\lrbrk{\frac{\pa\Omega_2}{\pa\pa_j\pa_k A_0}z^{\g_2}}\bigg)\d A_0\\
	&\qquad\qquad\qquad
	+\bigg(z^{\p^i} - \frac{\pa\Omega_1}{\pa A_i}z^{\g_1} + \pa_j\lrbrk{\frac{\pa\Omega_1}{\pa\pa_j A_i}z^{\g_1}}\\
	&\qquad\qquad\qquad\qquad
	- \frac{\pa\Omega_2}{\pa A_i}z^{\g_2}
	+ \pa_j\lrbrk{\frac{\pa\Omega_2}{\pa\pa_j A_i}z^{\g_2}}
	- \pa_j\pa_k\lrbrk{\frac{\pa\Omega_2}{\pa\pa_j\pa_k A_i}z^{\g_2}}
	\bigg)
	\d A_i\\
	&\qquad\qquad\qquad+(-z{^{A_0}}- z^{\g_1})\d\p^0
	+\bigg(-z^{A_i}
	- \frac{\pa\Omega_2}{\pa \p^i}z^{\g_2}
	+ \pa_j\lrbrk{\frac{\pa\Omega_2}{\pa\pa_j \p^i}z^{\g_2}}
	\bigg)\d\p^i\\
	&\qquad\qquad\qquad+\lrbrk{\frac{\pa\Omega_1}{\pa A_\m}z^{A_\m} + \frac{\pa\Omega_1}{\pa\pa_i A_j}\pa_iz^{A_j} + z^{\p^0}}\d\g_1\\
	&\qquad\qquad\qquad
	+\bigg(\frac{\pa\Omega_2}{\pa A_\m}z^{A_\m} + \frac{\pa\Omega_2}{\pa\pa_i A_\m}\pa_iz^{A_\m}
	+ \frac{\pa\Omega_2}{\pa\pa_i\pa_j A_\m}\pa_i\pa_j z^{A_\m}\\
	&\qquad\qquad\qquad\qquad
	+\frac{\pa\Omega_2}{\pa \p^i}z^{\p^i} + \frac{\pa\Omega_2}{\pa\pa_j \p^i}\pa_jz^{\p^i}
	\bigg)\d\g_2
	\Bigg),
	\end{split}
	\ee
	where partial derivatives for $\Omega_2$ are taken
	on $\Omega_2$ of the form
	\be
	\Omega_2(A_\m,\pa_i A_\m,\pa_i\pa_j A_\m,\p^i,\pa_j \p^i,g_{\m\n},\pa_\s g_{\m\n},\cdots,K).
	\ee
	We require that there is no further constraint.
	So let us demand that there is only a trivial solution to $i_z\cF = 0.$
	By eliminating $z^{\p^0}, z^{\g_1}, z^{A_i}$ from the coefficients of
	$\d\g_1, \d\p^0, \d\p^i,$ and substituting into the coefficient of $\d A_0,$
	we obtain
	\be\label{zg1-eqn}
	\begin{split}
	0&=-\lrbrk{\frac{\pa\Omega_2}{\pa\pa_j\pa_k A_0}
		+ \frac{\pa\Omega_1}{\pa\pa_{j} A_i}\frac{\pa\Omega_2}{\pa\pa_{k} \p^i}}\pa_j\pa_k z^{\g_2}\\
	&\qquad+
	\bigg(
	-\frac{\pa\Omega_1}{\pa A_i}\frac{\pa\Omega_2}{\pa \pa_j \p^i}
	+\frac{\pa\Omega_1}{\pa\pa_j A_i}\frac{\pa\Omega_2}{\pa\p^i}
	-\frac{\pa\Omega_1}{\pa \pa_i A_k}\pa_i\lrbrk{\frac{\pa\Omega_2}{\pa \pa_j \p^k}}
	-\frac{\pa\Omega_1}{\pa \pa_j A_k}\pa_i\lrbrk{\frac{\pa\Omega_2}{\pa \pa_i \p^k}}\\
	&\qquad\qquad+\frac{\pa\Omega_2}{\pa \pa_j A_0}
	-2\pa_k\lrbrk{\frac{\pa\Omega_2}{\pa \pa_j\pa_k A_0}}
	\bigg)\pa_j z^{\g_2}
	+
	z^{\g_2}\lrbrk{\int d^3\vec{y}\PB{\Omega_1}{\Omega_2(\vec{y})}},
	\end{split}
	\ee
	where we have imposed the condition \eq{condU0} on the Poisson bracket
	so that its integral reduces to
	\be
	\begin{split}
	\int d^3\vec{y}\PB{\Omega_1}{\Omega_2(\vec{y})}
	&=\frac{\pa\Omega_1}{\pa A_i}\frac{\pa\Omega_2}{\pa \p^i}
	- \frac{\pa\Omega_1}{\pa A_i}\pa_j\lrbrk{\frac{\pa\Omega_2}{\pa\pa_j \p^i} }
	+ \frac{\pa\Omega_1}{\pa \pa_i A_j}\pa_i\lrbrk{\frac{\pa\Omega_2}{\pa \p^j} }\\
	&\qquad- \frac{\pa\Omega_1}{\pa \pa_i A_j}\pa_i\pa_k\lrbrk{\frac{\pa\Omega_2}{\pa \pa_k\p^j} }
	- \frac{\pa\Omega_2}{\pa A_0}
	+ \pa_j\lrbrk{\frac{\pa\Omega_2}{\pa \pa_j A_0}}\\
	&\qquad- \pa_j\pa_k\lrbrk{\frac{\pa\Omega_2}{\pa \pa_j\pa_k A_0}}.
	\end{split}
	\ee
	In order for $\cF$ to possess no zero mode,
	we need to demand that the coefficients of $\pa_j z^{\g_2}$
	and of $\pa_j\pa_k z^{\g_2}$ in eq.\eq{zg1-eqn} vanish.
	At the same time, the coefficient of $z^{\g_2}$
	should be non-vanishing.
	This gives rise to the requirements
	\be\label{UVcond-4}
	\cC_2^{jk}\equiv\frac{\pa\Omega_2}{\pa\pa_j\pa_k A_0}
		+ \frac{\pa\Omega_1}{\pa\pa_{(j|} A_{i}}\frac{\pa\Omega_2}{\pa\pa_{|k)} \p^i} = 0.
	\ee	
	\be\label{UVcond-3}
	\begin{split}
	\cC_1^j
	&\equiv-\frac{\pa\Omega_1}{\pa A_i}\frac{\pa\Omega_2}{\pa \pa_j \p^i}
		+\frac{\pa\Omega_1}{\pa\pa_j A_i}\frac{\pa\Omega_2}{\pa\p^i}
		-\frac{\pa\Omega_1}{\pa \pa_i A_k}\pa_i\lrbrk{\frac{\pa\Omega_2}{\pa \pa_j \p^k}}\\
	&\qquad-\frac{\pa\Omega_1}{\pa \pa_j A_k}\pa_i\lrbrk{\frac{\pa\Omega_2}{\pa \pa_i \p^k}}
		+\frac{\pa\Omega_2}{\pa \pa_j A_0}
		-2\pa_k\lrbrk{\frac{\pa\Omega_2}{\pa \pa_j\pa_k A_0}}\\
	&=0,
	\end{split}
	\ee
	\be\label{UVcond-2}
	\int d^3\vec{y}\PB{\Omega_1}{\Omega_2(\vec{y})} \neq 0,
	\ee
	Imposing these requirements on eq.\eq{zg1-eqn}, one obtains $z^{\g_2} = 0.$
	After substituting this into eq.\eq{izcF-2} and requiring that the expression vanishes,
	one eventually sees that, without imposing any further conditions, the only solution
	to $i_z\cF = 0$ is $z=0.$

	So from the analysis, we see that if the theory \eq{LGP-UV}
	satisfies the conditions \eq{condU0}, \eq{UVcond-4}-\eq{UVcond-2},
	it has two constraints.
	Furthermore, the process guarantees that the constraints
	are of second-class. This is because the condition \eq{UVcond-2}
	requires that the Poisson brackets of $\Omega_1$ and $\Omega_2$ is non-vanishing.
	Now, since the theory has two second-class constraints
	just like the standard Proca theory,
	the counting of the degrees of freedom
	suggests that the theory has three degrees of freedom
	as required.

	An alternative way to see that, under
	the conditions \eq{condU0}, \eq{UVcond-4}-\eq{UVcond-2},
	the theory possesses
	two second-class constraints is by using the criteria of \cite{Rodrigues:2018ioe}.
	That is, since each zero mode of $\cF$ in any step
	leads to an independent constraint, there is no gauge symmetry
	in the theory, and hence all of the constraints
	found are of second-class.
	On the other hand, if we suppose that some of the conditions \eq{UVcond-4}-\eq{UVcond-2}
	are not satisfied,
	then there exists further zero modes.
	In case these zero modes do not lead to any new constraint,
	the criteria of \cite{Rodrigues:2018ioe} suggests that
	some of the constraints already obtained are of first-class.
	So the number of degrees of freedom is less than three.
	Alternatively, if the zero modes do lead to new constraints,
	then the theory possesses at least three constraints.
	So in the case, the number of degrees of freedom is also less than three.

	In fact, the conditions \eq{condU0} and \eq{UVcond-4}-\eq{UVcond-3}
	are implied by diffeomorphism invariance.
	We have shown that this is the case for the condition \eq{condU0}.
	So let us show this for the conditions \eq{UVcond-4}-\eq{UVcond-3}.
	
	\subsection{Triviality of $\cC_2^{ij} = 0$}

	The condition \eq{UVcond-4} can easily be shown
	to be automatically satisfied following the diffeomorphism invariance requirement.
	For this, let us first express $\Omega_2$ in phase space
	and keep only terms containing $\pa_i\pa_j A_0$
	and $\pa_i\p^i.$
	So only the relevant terms are
	\be\label{O2C2}
	\begin{split}
	\Omega_2
	&= -\pa_i\lrbrk{\frac{\pa\cT}{\pa\pa_i A_0}} - \pa_j\L_i\frac{\pa U}{\pa\pa_j A_i} + (\textrm{terms free from\ $\pa_i\pa_j A_0$ and $\pa_i\p^i$})\\
	&= -\frac{\pa^2\cT}{\pa\pa_i A_0\pa\pa_j A_0}\pa_i\pa_j A_0
	-\lrbrk{\frac{\pa^2\cT}{\pa\L_i\pa\pa_j A_0}
	+ \frac{\pa U}{\pa\pa_j A_i}}\pa_j\L_i\\
	&\qquad+ (\textrm{terms free from\ $\pa_i\pa_j A_0$ and $\pa_i\p^i$})\\
	&= \frac{\pa^2\cT}{\pa\L_i\pa\L_j}\pa_j\pa_k A_0\lrbrk{-\d_i^k + \frac{\pa\L_i}{\pa\pa_k A_0}}
	+\frac{\pa^2\cT}{\pa\L_i\pa\L_j}\frac{\pa\L_i}{\pa\p^m}\pa_j\p^m\\
	&\qquad+ (\textrm{terms free from\ $\pa_i\pa_j A_0$ and $\pa_i\p^i$}),
	\end{split}
	\ee
	where in the third step,
	we used eq.\eq{TTU}-\eq{condTV} with replacement $\pa_0A_i\to\L_i.$
	The expression for $\Omega_2$ in eq.\eq{O2C2} can be further simplified.
	For this, let us compute partial derivatives of $\L_i$ with respect to $\p^i$
	and to $\pa_i A_0,$ we consider derivatives of $\p^i = \pa\cT/\pa\L_i$
	with respect to phase space variables. This gives
	\be\label{paLpapi}
	\frac{\pa^2\cT}{\pa\L_i\pa\L_m}\frac{\pa\L_i}{\pa\p^m} = \d^i_m,
	\ee
	\be\label{paLpajA0}
	0 = \frac{\pa^2\cT}{\pa\L_i\pa\pa_j A_0} + \frac{\pa^2\cT}{\pa\L_i\pa\L_k}\frac{\pa\L_k}{\pa\pa_j A_0}.
	\ee	
	By using these equations and eq.\eq{TTU}-\eq{condTV} again,
	we obtain
	\be
	\Omega_2\supset\frac{\pa U}{\pa_j A_i}\pa_i\pa_j A_0 + \pa_i\p^i.
	\ee
	By substituting this into eq.\eq{UVcond-4},
	it can be seen that this is automatically satisfied.

	So only the conditions \eq{UVcond-2}-\eq{UVcond-3}
	are actually required for the theory to have
	three degrees of freedom. Other conditions
	are already satisfied thanks to the requirement
	of diffeomorphism invariance.	
	
	So from the analysis, we see that if the theory \eq{LGP-UV}
	satisfies the conditions \eq{UVcond-2}-\eq{UVcond-3},
	then it has two constraints.
	Furthermore, the process guarantees that the constraints
	are of second-class. This is because the condition \eq{UVcond-2}
	requires that the Poisson brackets of $\Omega_1$ and $\Omega_2$ vanish.
	Now, since the theory has two second-class constraints
	just like the standard Proca theory,
	the counting of the degrees of freedom
	suggests that the theory has three degrees of freedom
	as required.

	Combining with the result we obtained previously,
	we may conclude the finding so far as follows.
	The requirement of diffeomorphism invariance
	demands the theory to have more than one constraint.
	Next, if there is no zero mode of the symplectic $2-$form
	at the second iteration, then the theory is guaranteed
	to have two constraints with both being of second-class.

	\subsection{Triviality of $\cC_1^j = 0$}
	Let us now show that the condition \eq{UVcond-3}
	is automatically satisfied by diffeomorphism invariance requirement.
	To rewrite the condition \eq{UVcond-3},
	we need to first express $\Omega_1$ and $\Omega_2$
	in terms of $T, U$ and their derivatives.
	Expressing $\Omega_1 = \p^0 - U$ is a simple task.
	So we need to express $\Omega_2.$
	In the previous subsection, we have already computed
	\be
	\frac{\pa\Omega_2}{\pa\pa_i \p^j},\qquad
	\frac{\pa\Omega_2}{\pa\pa_i\pa_j A_0}.
	\ee
	So we are left to compute
	\be
	\frac{\pa\Omega_2}{\pa\p^i},\qquad
	\frac{\pa\Omega_2}{\pa\pa_j A_0}.
	\ee
		
	Let us revisit eq.\eq{constraint2}.
	It can be rewritten as
	\be\label{O2-torewrite}
	\Omega_2 = \frac{\pa\cT}{\pa A_0} - \pa_j\lrbrk{\frac{\pa\cT}{\pa\pa_j A_0} + \frac{\pa U}{\pa\pa_j A_i}\L_i}
	-\lrbrk{\frac{\pa U}{\pa A_i} - \pa_j\lrbrk{\frac{\pa U}{\pa\pa_j A_i}}}\L_i.
	\ee
	Consider $\pa\Omega_2/\pa\p^i.$
	It can be shown that the second term does not depend on $\p^i.$
	To see this, we consider a diffeomorphism condition \eq{TUfull}.
	Changing the variables to phase space ones gives
	\be\label{pake0cond}
	\begin{split}
	\frac{\pa\cT}{\pa\pa_k A_0} + \frac{\pa U}{\pa_k A_i}\L_i
	&=
	-\p^k - \frac{\pa U}{\pa A_k} A_0 - \frac{\pa U}{\pa\pa_i A_k}\pa_i A_0\\
	&\qquad-2\frac{\pa U}{\pa g_{\m k}}g_{\m0} - \frac{\pa U}{\pa\pa_k g_{\r\s}}\pa_0 g_{\r\s}
	-2\frac{\pa U}{\pa\pa_\m g_{\r k}}\pa_\m g_{\r 0} + \cdots,
	\end{split}
	\ee
	where $\cdots$ are terms involving partial derivatives of $U$
	with respect to higher order derivatives of $g_{\m\n}$
	and to external tensor fields and their derivatives.
	Since $U$ does not depend on $\p^i,$
	it can easily be seen that after applying the partial derivative $\pa_k$
	on the above equation, the resulting expression does not depend on $\p^k.$
	Similarly, it can easily be seen that the coefficient
	of $\L_i$ in the last term of eq.\eq{O2-torewrite}
	do not depend on $\p^i.$
	So we have
	\be
	\frac{\pa\Omega_2}{\pa\p^i}
	= \lrbrk{\frac{\pa^2\cT}{\pa A_0\pa\L_k}
	-\frac{\pa U}{\pa A_k} + \pa_j\lrbrk{\frac{\pa U}{\pa\pa_j A_k}}}\frac{\pa\L_k}{\pa\p^i}.
	\ee
	Next, let us compute $\pa\Omega_2/\pa\pa_j A_0.$
	We obtain
	\be
	\begin{split}
	\frac{\pa\Omega_2}{\pa\pa_j A_0}
	&=\frac{\pa^2\cT}{\pa A_0\pa\pa_j A_0}
	+\lrbrk{\frac{\pa^2\cT}{\pa A_0\pa\L_k}
		-\frac{\pa U}{\pa A_k} + \pa_m\lrbrk{\frac{\pa U}{\pa\pa_m A_k}}}\frac{\pa\L_k}{\pa \pa_j A_0}
	+\frac{\pa U}{\pa A_j}\\
	&\qquad+\frac{\pa^2 U}{\pa_j A_i\pa A_0}\L_i\\
	&\qquad+\frac{\pa^2 U}{\pa A_j \pa A_0}A_0
	+\pa_k\lrbrk{\frac{\pa U}{\pa\pa_j A_k}}
	+\frac{\pa^2 U}{\pa \pa_k A_j \pa A_0}\pa_k A_0\\
	&\qquad
	+2\frac{\pa^2 U}{\pa g_{\m j}\pa A_0}g_{\m 0}
	+\frac{\pa^2 U}{\pa\pa_j g_{\r\s}\pa A_0}\pa_0 g_{\r\s}
	+2\frac{\pa^2 U}{\pa\pa_\m g_{\r j}\pa A_0}\pa_\m g_{\r 0}\\
	&\qquad+\cdots\\
	&=
	\lrbrk{\frac{\pa^2\cT}{\pa A_0\pa\L_k}
		-\frac{\pa U}{\pa A_k} + \pa_m\lrbrk{\frac{\pa U}{\pa\pa_m A_k}}}\frac{\pa\L_k}{\pa \pa_j A_0}\\
	&\qquad
	+\pa_k\lrbrk{\frac{\pa U}{\pa\pa_j A_k}}
	-\frac{\pa^2\cT}{\pa\L_k \pa A_0},
	\end{split}
	\ee
	where the first step is obtained by first applying eq.\eq{pake0cond}
	and then taking derivative, and in the second step, we apply
	eq.\eq{pake0cond} again.
	
	With these ingredients, we see that $\cC_1^j$
	is given by
	\be\label{cCrewrite}
	\begin{split}
	\cC_1^j
	&=
	\lrbrk{-\d_k^j+\frac{\pa\L_k}{\pa \pa_j A_0}-\frac{\pa U}{\pa\pa_j A_m}\frac{\pa\L_k}{\pa\p^m}}
	\lrbrk{\frac{\pa^2\cT}{\pa A_0\pa\L_k}
		-\frac{\pa U}{\pa A_k} + \pa_i\lrbrk{\frac{\pa U}{\pa\pa_i A_k}}}.		
	\end{split}
	\ee	
	It can be further simplified.
	For this, let us consider
	\be\label{cTTfk}
	\begin{split}
	\frac{\pa^2\cT}{\pa\L_k\pa\L_n}\bigg(-\d_k^j+\frac{\pa\L_k}{\pa \pa_j A_0}-&\frac{\pa U}{\pa\pa_j A_m}\frac{\pa\L_k}{\pa\p^m}\bigg)\\
	&=\lrbrk{-\frac{\pa^2\cT}{\pa\L_j\pa\L_n}-\frac{\pa^2\cT}{\pa\L_n{\pa \pa_j A_0}}-\frac{\pa U}{\pa\pa_j A_n}}\\
	&=0,
	\end{split}
	\ee
	where in the first step, we use eq.\eq{paLpapi}-\eq{paLpajA0},
	and in the second step, we used eq.\eq{condTV} with change of variables $\pa_0 A_i\to\L_i.$
	Next, the expression $\pa^2\cT/\pa\L_k\pa\L_n$ is always invertible.
	To see this, we note from a requirement in this paper that
	\be
	\det\lrbrk{\frac{\pa^2\cL}{\pa\dot A_i\pa\dot A_j}} \neq 0.
	\ee
	By noting that $U$ does not depend on $\dot A_i$
	and by translating this equation to phase space,
	we obtain the condition that
	\be
	\det\lrbrk{\frac{\pa^2\cT}{\pa\L_i\pa\L_j}} \neq 0.
	\ee		
	With this condition, eq.\eq{cTTfk} then reduces to
	\be
	-\d_k^j+\frac{\pa\L_k}{\pa \pa_j A_0}-\frac{\pa U}{\pa\pa_j A_m}\frac{\pa\L_k}{\pa\p^m} = 0.
	\ee
	This further simplifies \eq{cCrewrite} to
	\be\label{cCrewrite-2}
	\begin{split}
	\cC_1^j
	&=0.		
	\end{split}
	\ee
	That is, the condition \eq{UVcond-3} is automatically satisfied by diffeomorphism invariance.

	\setcounter{equation}0
	\section{Example cases}\label{sec:example}
	In the previous section, we have shown that
	a vector theory which satisfies the conditions \eq{Hzeromu}
	are almost guaranteed to have three propagating degrees of freedom.
	This is largely due to diffeomorphism invariance requirements.	
	
	Suppose that one proceeds directly to count the number of propagating degrees of freedom
	of a generalised Proca theory using the Faddeev-Jackiw method.
	If the theory is diffeomorphic invariance and satisfied \eq{Hzeromu},
	one is going to obtain the following results:
	\begin{enumerate}
	  \item[(I)] there exists at least two constraints, and that
	  \item[(II)] it is likely that there are only two constraints, and both constraints are of second-class.
	\end{enumerate}
	We learned from the previous section that these two results
	for any diffeomorphic invariance generalised Proca theory
	which satisfies the conditions \eq{Hzeromu}
	are due to diffeomorphism invariance requirements.
	In particular, the result (I) is due to diffeomorphism invariance requirements.
	Specifically eq.\eq{condU0} trivialises eq.\eq{zA0-eqn}.
	This in turn points out the existence of the second constraint.
	As for the result (II),
	diffeomorphism invariance requirements trivialises eq.\eq{UVcond-4}-\eq{UVcond-3}.
	If the condition \eq{UVcond-2} is also satisfied,
	the theory is guaranteed to have
	two constraints, and both constraints are of second-class.
	Although the condition \eq{UVcond-2} is not trivialised by diffeomorphism invariance requirements,
	it is easy to be satisfied.
	This is in the sense that the condition \eq{UVcond-2}
	demands a complicated expression not to trivially vanishes.
	
	In this section, we demonstrate this by showing simple examples
	and show by direct calculation (without using the results from the previous section)
	of the Faddeev-Jackiw method
	that these examples satisfy both the results (I) and (II).
	
	\subsection{A special case in flat spacetime}	
	Let us consider a special case of eq.\eq{L2toL6genProca}
	in flat spacetime.
	We consider the case where $\b_6 = 0, G_2 = G_2(A_\m A^\m), \tilde G_5 = 0.$
	Explicitly, in this example, we consider the Lagrangian
	\be\label{LgenProca}
	{\cal L}_{GP}=-\frac{1}{4} F_{\m\n}F^{\m\n}+\sum_{n=2}^{5} \a_{n}{\cal L}_n,
	\ee
	where
	$\a_{n}, n=2,3,4,5$ are arbitrary constants and ${\cal L}_n, n=2,3,4,5$
	are self-interactions of the gauge field given by
	\bea
	{\cal L}_{2}&=&f_{2}(A^2),
\label{L2genProca}\nonumber\\
	{\cal L}_{3}&=&f_{3}(A^2)\pa \cdot A,
\label{L3genProca}\nonumber\\
	{\cal L}_{4}&=&f_{4}(A^2)[(\pa \cdot A)^{2} - \pa_{\r} A_{\s}\pa^{\s} A^{\r}],
\label{L4genProca}\nonumber\\
	{\cal L}_{5}&=&f_{5}(A^2)\big[(\pa \cdot A)^{3} -3 (\pa \cdot A) \pa_{\r} A_{\s}\pa^{\s} A^{\r}
	+2\pa_{\r} A_{\s}\pa^{\g} A^{\r} \pa^{\s} A_{\g}\big],
\label{L5genProca}
\eea
	where $A^{2}\equiv A_{\m}A^{\m}$, $\pa \cdot A\equiv\pa_{\m} A^{\m}.$

	We have directly followed Faddeev-Jackiw process
	and obtained two constraints. Furthermore, each zero mode
	of symplectic $2-$form in each iteration leads to a constraint.
	So the theory has three propagating degrees of freedom as expected.
	This by itself suggests that the theory eq.\eq{LgenProca}
	satisfies the conditions \eq{condU0} and \eq{UVcond-4}-\eq{UVcond-2}.
	
	As a cross check of these conditions,
	we have substituted the constraints
	\be\label{primary-constraint-genProca}
	\Omega_{1}\equiv\p^0 + \a_3 f_3(A^2) + 2\a_4 f_4(A^2)\vec{\na}\cdot\vec{A} + 3\a_5 f_5(A^2)((\vec{\na}\cdot\vec{A})^2 - \pa_iA_j\pa_jA_i),
	\ee	
	\be
	\begin{split}
	\Omega_{2}
	&=\vec{\nabla }\cdot \vec{\pi }-2 \alpha _2f_2'(A^2)A_0
	+2\alpha_3 f_3'(A^2)\left(\vec{A}\cdot (\vec{\pi } + \vec{\nabla }A_0) -A_0\vec{\nabla }\cdot \vec{A} \right) \\
	&\qquad+\alpha _4\bigg(4f_4'(A^2)\Big(2A^{[i}(\p_i + \pa_iA_0)\pa^{j]}A_j
	- A_\m\partial _jA^\m\partial _jA_0 \\
	&\qquad\qquad\qquad\qquad\qquad + A_0\partial^{[i}A_j\partial^{j]}A_i\Big)-2 f_4(A^2) \nabla^2A_0\bigg) \\
	&\qquad-4 \alpha _3 \alpha _4 f_3'(A^2) f_4(A^2)\vec{A}\cdot \vec{\nabla }A_0
	+16\alpha _4^2f_4(A^2) f_4'(A^2) A^{[i}\partial _jA_0\partial^{j]}A_i \\
	&\qquad+12\alpha _5\bigg(f_5'(A^2)\Big(3 A_i (\p^{[i}+ \pa^{[i}A_0) \pa_j A^j\pa_kA^{k]}
	+2A_\m\partial_jA^\m\partial^{[j}A_k\partial^{k]}A_0\\
	&\qquad\qquad\qquad\qquad\qquad-A_0\pa^{[i}A_i\pa^{j}A_j\pa^{k]}A_k\Big)
	+ f_5(A^2) \partial^{[i}\lrbrk{\pa_j A_0\partial_iA^{j]}} \bigg) \\
	&\qquad+24\alpha _3 \alpha _5f_3'(A^2) f_5(A^2) A_i\partial^{[j}A_0\partial^{i]}A_j \\
	&\qquad-24\alpha _4 \alpha _5\bigg(
	4f_4'(A^2) f_5(A^2) A^{[i}\pa^{k]}A_k\pa^{[i}A_0\pa^{j]}A_j\\
	&\qquad\qquad\qquad\qquad\qquad+3 f_4(A^2) f_5'(A^2) A_i\pa^{[i}A_0\pa_j A^{j}\pa_k A^{k]}\bigg)\\
	&\qquad-432\alpha _5^2f_5(A^2) f_5'(A^2)A^{[i}\pa^j A_j\pa^{k]}A_k \pa_{[i|}A_0\pa_{|l]}A^l
	\end{split}
	\ee	
	into the conditions \eq{condU0} and \eq{UVcond-4}-\eq{UVcond-2} and see that they are indeed satisfied.
	
	\subsection{$U=0$}	
	Let us now turn to the case where $U=0,$
	and the vector field couples to the metric as well as other background fields.
	This example is in fact simple enough to be directly demonstrated.
	
	In this case, the Lagrangian is
	\be\label{cL-U0}
	\cL = \cL(A_\m,\pa_\m A_\n, g_{\m\n},\pa_\r g_{\m\n},\cdots,K).
	\ee
	If we trade $\pa_\m A_\n$ for $F_{\m\n}$ and $S_{\m\n}\equiv\na_\m A_\n + \na_\n A_\m,$
	it can be seen that the Lagrangian should not depend on $S_{\m\n},$
	otherwise $S_{00} \supset 2\dot A_0$ would appear in the Lagrangian,
	and hence does not agree with our requirement in this example.
	Therefore,
	\be
	\cL = \cL(A_\m,F_{ij}, F_{0i}, g_{\m\n},\pa_\r g_{\m\n},\cdots,K).
	\ee
	Furthermore, since $U=0$ eq.\eq{LGP-UV} reduces to
	\be\label{cL-T}
	\cL = T.
	\ee
	Computing conjugate momenta, we arrive at the primary constraint
	\be
	\Omega_1 = \p^0,
	\ee
	and the conditions which can be used to trade between $\p^i$ and $F_{0i}:$
	\be\label{pi-F0i}
	\p^i = \frac{\pa T}{\pa F_{0i}}.
	\ee
	The inversion of this equation is of the form
	\be\label{F0i}
	F_{0i} = \tilde{\L}_i(A_\m,F_{kl},\p^k, g_{\m\n},\pa_\r g_{\m\n},\cdots,K).
	\ee
	The results given in the previous section
	relies on the quantities $\L_i$ instead of $\tilde{\L}_i.$
	So when needed to compare with them, we may recover $\L_i$ from
	\be
	\L_i = \tilde{\L}_i + \pa_i A_0.
	\ee
	Next, by substituting eq.\eq{F0i}
	into eq.\eq{cL-T}, we obtain
	\be\label{cTT}
	\tilde{\cT} = \tilde{\cT}(A_\m,F_{ij},\tilde{\L}_i, g_{\m\n},\pa_\r g_{\m\n},\cdots,K).
	\ee
	Furthermore, eq.\eq{pi-F0i} in phase space is
	\be
	\p^i = \frac{\pa \tilde{\cT}}{\pa \tilde{\L}_i}.
	\ee
	The secondary constraint can then be worked out to be
	\be
	\Omega_2 = \frac{\pa\tilde{\cT}}{\pa A_0} + \pa_i\p^i.
	\ee
	
	It can be seen that, as expected, this theory satisfies
	the conditions \eq{condU0}, \eq{UVcond-4}-\eq{UVcond-3}.
	The most non-trivial check is on the condition \eq{UVcond-3}.
	This reduces to
	\be\label{condUeq0}
	\frac{\pa\Omega_2}{\pa\pa_j A_0} = 0.
	\ee
	This condition is trivially satisfied since
	after expressing $\tilde{\cT}$ in phase space variables
	(by substituting $\tilde\L_i$ from eq.\eq{F0i} into eq.\eq{cTT}),
	we see that $\tilde{\cT}$ do not depend on $\pa_i A_0.$

	As for the condition \eq{UVcond-2},
	some theories in this example might not satisfy this.
	We see that it reduces to
	\be
	\frac{\pa^2\tilde{\cT}}{\pa A_0^2} \neq 0,
	\ee
	or after changing the variables to configuration space,
	\be
	\frac{\pa^2 T}{\pa A_0^2} \neq 0.
	\ee
	This is the only condition that the Lagrangian \eq{cL-U0}
	has to satisfy in order for the vector field to have three propagating degrees of freedom.	

	As an example, a generalised Proca theory
	with Lagrangian of the form \eq{cL-U0}
	is
	\be
	\cL = -\sqrt{-g}\ove{4}F_{\m\n}F^{\m\n} + \sqrt{-g} G_2(A_\m, F_{\m\n}, g_{\m\n}).
	\ee
	This theory has three propagating degrees of freedom if
	\be
	\frac{\pa^2 G_2}{\pa A_0^2} \neq 0.
	\ee

	\setcounter{equation}0	
	\section{On Faddeev-Jackiw brackets}\label{sec:FJB}
	For completeness, let us discuss
	the process to compute Faddeev-Jackiw brackets
	of diffeomorphism invariance generalised Proca theories
	satisfying the requirement \eq{Hzeromu}.
	Furthermore, we focus on the cases which also satisfy the condition \eq{UVcond-2}.
	
	In section \ref{sec:cond-constraint}, the Faddeev-Jackiw constrained analysis
	was performed on these theories.
	At the final iteration, we obtain the symplectic $2-$form
	as shown in eq.\eq{CF-2ndit}.
	Let us write this in matrix form.
	For this, we first re-express eq.\eq{CF-2ndit}
	in the form
	\be\label{cF-genProca-fin}
	\cF = \hlf\int d^3\vec{x}\int d^3\vec{y} \cF_{\xi^I\xi^J}(\vec{x},\vec{y})\d\xi^I(\vec{x})\w\d\xi^J(\vec{y}),
	\ee
	where $\xi^I = (A_\m,\p^\n,\g_1,\g_2),$
	and $\xi^J = (A_\r,\p^\s,\g_1,\g_2).$
	The quantities $\cF_{\xi^I\xi^J}(\vec{x},\vec{y})$ appearing
	in eq.\eq{cF-genProca-fin} are elements of the matrix
	\be\label{cF-genProca-fin-matrix}
	\cF(\vec{x},\vec{y})
	=
	\begin{pmatrix}
	\mathbf{A}(\vec{x},\vec{y}) & \mathbf{B}(\vec{x},\vec{y})\\
	\mathbf{C}(\vec{x},\vec{y}) & \mathbf{D}(\vec{x},\vec{y})
	\end{pmatrix}
	,
	\ee
	where $\mathbf{A}, \mathbf{B}, \mathbf{C},$ and $\mathbf{D}$
	are block matrices given by
	\be
	\mathbf{A}(\vec{x},\vec{y})
	=
	\begin{pmatrix}
		0 & -\d^\m{}_\s\\
		\d_\n{}^\r & 0
	\end{pmatrix}
	\d^{(3)}(\vec{x}-\vec{y}),
	\ee
	\be
	\mathbf{B}(\vec{x},\vec{y})
	=
	\begin{pmatrix}
	F^\m(x)+ \d^\m_j G^{ji}(y)\pa_{y^i} & M^\m(x)+ N^{\m i}(y)\pa_{y^i} + P^{\m ij}(y)\pa_{y^i}\pa_{y^j} \\
	\d_\n{}^0 & \d_{\n}^i(Q_i(x)+\d_i^j\pa_{y^j})
	\end{pmatrix}
	\d^{(3)}(\vec{x}-\vec{y}),
	\ee
	\be
	\mathbf{C}(\vec{x},\vec{y})
	=
	\begin{pmatrix}
	-F^\r(x) - \d^\r_j G^{j i}(x)\pa_{x^i}& -\d^0{}_\s\\
	-M^\r(x)- N^{\r i}(x)\pa_{x^i} - P^{\r ij}(x)\pa_{x^i}\pa_{x^j} & -\d^i{}_\s(Q_i(x)+\d_i^j\pa_{x^j})
	\end{pmatrix}
	\d^{(3)}(\vec{x}-\vec{y}),
	\ee
	\be
	\mathbf{D}(\vec{x},\vec{y})
	=
	\begin{pmatrix}
		0 & 0\\
		0 & 0
	\end{pmatrix}
	,
	\ee
	with
	\be
	F^\m \equiv \frac{\pa\F_{1}}{\pa A_\m},\qquad
	G^{ji}\equiv \frac{\pa\F_{1}}{\pa \pa_i A_j},
	\ee
	\be
	M^\m \equiv \frac{\pa\F_{2}}{\pa A_\m},\qquad
	N^{\m i}\equiv \frac{\pa\F_{2}}{\pa \pa_i A_\m},\qquad
	P^{\m ij}\equiv \frac{\pa\F_{2}}{\pa \pa_i\pa_j A_\m},
	\ee
	\be
	Q_i \equiv \frac{\pa\F_{2}}{\pa \p^i}.
	\ee
	Note that, we have used the fact that
	\be
	\frac{\pa\F_1}{\pa \p^\m} = \d_\m^0,
	\qquad
	\frac{\pa\F_2}{\pa \pa_j\p^i} = \d_i^j,
	\ee
	which can easily be obtained from eq.\eq{constraint1}
	and eq.\eq{constraint2}.
	
	The matrix \eq{cF-genProca-fin-matrix} is already in the block form.
	So its inverse can be obtained
	from the formula
	\be\label{Finv}
	\cF^{-1}
	=
	\begin{pmatrix}
	\mathbf{A}^{-1} + \mathbf{A}^{-1}\mathbf{B}(\mathbf{D}-\mathbf{C}\mathbf{A}^{-1}\mathbf{B})^{-1}\mathbf{C}\mathbf{A}^{-1} & -\mathbf{A}^{-1}\mathbf{B}(\mathbf{D}-\mathbf{C}\mathbf{A}^{-1}\mathbf{B})^{-1}\\
	-(\mathbf{D}-\mathbf{C}\mathbf{A}^{-1}\mathbf{B})^{-1}\mathbf{C}\mathbf{A}^{-1} & (\mathbf{D}-\mathbf{C}\mathbf{A}^{-1}\mathbf{B})^{-1}
	\end{pmatrix}
	.
	\ee
	The least straightforward step in the calculation
	of the matrix $\cF^{-1}$ is to compute $(\mathbf{D}-\mathbf{C}\mathbf{A}^{-1}\mathbf{B})^{-1}.$
	For this, one first needs to compute $\mathbf{D}-\mathbf{C}\mathbf{A}^{-1}\mathbf{B},$
	and then find the inverse.
	Let us present $\mathbf{D}-\mathbf{C}\mathbf{A}^{-1}\mathbf{B}$
	without yet imposing the conditions \eq{UVcond-4}-\eq{UVcond-3}.
	We have
	\be\label{dcab-init}
	(\mathbf{D}-\mathbf{C}\mathbf{A}^{-1}\mathbf{B})(\vec{x},\vec{y})
	=
	\begin{pmatrix}
	(\mathbf{D}-\mathbf{C}\mathbf{A}^{-1}\mathbf{B})_{11}(\vec{x},\vec{y}) & (\mathbf{D}-\mathbf{C}\mathbf{A}^{-1}\mathbf{B})_{12}(\vec{x},\vec{y}) \\
	(\mathbf{D}-\mathbf{C}\mathbf{A}^{-1}\mathbf{B})_{21}(\vec{x},\vec{y}) & (\mathbf{D}-\mathbf{C}\mathbf{A}^{-1}\mathbf{B})_{22}(\vec{x},\vec{y})
	\end{pmatrix}
	,
	\ee
	where
	\be\label{dcab-11}
	(\mathbf{D}-\mathbf{C}\mathbf{A}^{-1}\mathbf{B})_{11}(\vec{x},\vec{y}) = 0,
	\ee
	\be\label{dcab-12}
	\begin{split}
	(\mathbf{D}-\mathbf{C}\mathbf{A}^{-1}\mathbf{B})_{12}(\vec{x},\vec{y})
	&= (-\cC_2^{ij}(\vec{x})\pa_{x^i}\pa_{x^j}+\cC^i_1(\vec{x})\pa_{x^i}-\tilde{\cM}^0(\vec{x}))\d^{(3)}(\vec{x}-\vec{y})\\
	&=
	-(\mathbf{D}-\mathbf{C}\mathbf{A}^{-1}\mathbf{B})_{21}(\vec{y},\vec{x}),
	\end{split}
	\ee
	\be\label{dcab-22}
	\begin{split}
	(\mathbf{D}-\mathbf{C}&\mathbf{A}^{-1}\mathbf{B})_{22}(\vec{x},\vec{y})\\
	&= (-(P^{ijk}(\vec{x})+P^{ijk}(\vec{y}))\pa_{x^i}\pa_{x^j}\pa_{x^k}-(\tilde{\cM}^i(\vec{x})+\tilde{\cM}^i(\vec{y}))\pa_{x^i})\d^{(3)}(\vec{x}-\vec{y}),
	\end{split}
	\ee
	with
	\be\label{cMt0}
	\tilde{\cM}^0 = M^0 - G^{ji}\pa_iQ_j - \pa_i N^{0i} + \pa_{ij}P^{0ij}- F^i Q_i,
	\ee
	\be\label{cMti}
	\tilde{\cM^i} = M^i -\pa_j N^{(ij)} - N_i^\s Q_\s + \pa_j P^{\s ij}Q_\s - P^{\s ij}\pa_jQ_\s.
	\ee
	
	We see that there is the presence of $\cC^{ij}_2, \cC^i_1$
	in the $(12)-$ and $(21)-$components of $(\mathbf{D}-\mathbf{C}\mathbf{A}^{-1}\mathbf{B}).$
	In order to see the significance of the conditions \eq{UVcond-4}-\eq{UVcond-3},
	which follows from diffeomorphism invariance requirements,
	let us first suppose that these conditions are not satisfied.
	When $\cC^{ij}_2$ and $\cC^i_1$ do not simultaneously vanish,
	the inverse of $(\mathbf{D}-\mathbf{C}\mathbf{A}^{-1}\mathbf{B})$
	would contain infinite order derivative on delta function.
	For example, the $(12)$-component of
	$(\mathbf{D}-\mathbf{C}\mathbf{A}^{-1}\mathbf{B})^{-1}$
	is the inverse of the expression of the form
	\be\label{padelta2}
	(a^{ij}(\vec{x})\pa_{x^i}\pa_{x^j} + b^{i}(\vec{x})\pa_{x^i} + c(\vec{x}))\d^{(3)}(\vec{x}-\vec{y}).
	\ee
	The inverse of \eq{padelta2} cannot be expressed
	using linear combinations of finite terms of the form $\pa_{x^{i_1}\cdots x^{i_n}}\d^{(3)}(\vec{x}-\vec{y}).$
	To further illustrate the point,
	consider a one-dimensional toy example
	\be\label{thetoy}
	\d(x-y)+\pa_x\d(x-y).
	\ee
	It can easily be worked out that the inverse of \eq{thetoy}
	is
	\be\label{thetoy-inverse}
	\sum_{r=0}^\infty (-1)^r\pa_x^r\d(x-y).
	\ee
	So it can be expected that
	the inverse of the expression of the form \eq{padelta2}
	would surely contain infinite order derivative on delta function.
	On the other hand, if we now impose the conditions \eq{UVcond-4}-\eq{UVcond-3},
	then the matrix \eq{dcab-init} reduces to
	\be
	(\mathbf{D}-\mathbf{C}\mathbf{A}^{-1}\mathbf{B})(\vec{x},\vec{y})
	=
	\begin{pmatrix}
	0 & -\tilde{\cM}^0(\vec{x})\\
	\tilde{\cM}^0(\vec{y}) & -(\tilde{\cM}^i(\vec{x})+\tilde{\cM}^i(\vec{y}))\pa_{x^i}
	\end{pmatrix}
	\d^{(3)}(\vec{x}-\vec{y}).
	\ee
	Its inverse now contains finite linear combinations
	of the expressions of the form $\pa_{x^{i_1}\cdots x^{i_n}}\d^{(3)}(\vec{x}-\vec{y}).$
	More explicitly,
	\be\label{dcab-fin}
	\begin{split}
	(&\mathbf{D}-\mathbf{C}\mathbf{A}^{-1}\mathbf{B})^{-1}(\vec{x},\vec{y})\\
	&=
	\begin{pmatrix}
		-\!\lrbrk{W^{ijk}(\vec{x})\!+\!W^{ijk}(\vec{y})}\!\pa_{x^i}\pa_{x^j}\pa_{x^k}\!-\!\lrbrk{W^i(\vec{x})+W^i(\vec{y})}\!\pa_{x^i} & W^0(\vec{x})\\
		-W^0(\vec{x}) & 0
	\end{pmatrix}
	\d^{(3)}(\vec{x}-\vec{y}),
	\end{split}
	\ee
	where
	\be
	W^{ijk}\equiv\frac{P^{ijk}}{\tilde{\cM}^0},\qquad
	W^i\equiv\frac{\tilde{\cM}^i}{(\tilde{\cM}^0)^2} + 3\frac{\pa_k\tilde{\cM}^0}{(\tilde{\cM}^0)^2}\pa_j\lrbrk{\frac{P^{(ijk)}}{\tilde{\cM}^0}},\qquad
	W^0\equiv\ove{\tilde{\cM}^0}.
	\ee
	One may then substitute eq.\eq{dcab-fin}
	into eq.\eq{Finv} to obtain $\cF^{-1}$
	and hence the Faddeev-Jackiw bracket.

	\setcounter{equation}0	
	\section{Conclusion}\label{sec:conclusion}
	In this paper, we have shown that
	a generalised Proca theory
	coupled to any background field
	and satisfying eq.\eq{Hzeromu}
	is likely to have three propagating degrees of freedom
	in the vector sector provided that the theory is diffeomorphism invariant.
	By using Faddeev-Jackiw analysis,
	we have arrived at several conditions that the theory should satisfy
	in order to obtain three propagating degrees of freedom.
	It turns out that diffeomorphism invariance trivialises
	almost all the conditions except for the condition \eq{UVcond-2}.
	This condition demands a complicated combination of terms
	to not be trivially zero. So this condition can easily be fulfilled.
	
	For completeness, we have also investigated
	on how diffeomorphism invariance helps
	in simplifying Faddeev-Jackiw brackets.
	It turns out that diffeomorphism invariance
	requires that Faddeev-Jackiw brackets
	should be expressible using linear combinations
	of finite terms of partial derivatives on Dirac delta function.
	It would be interesting to investigate the implication of this result in details later.
	
	Although the analysis in this paper is given in four-dimensional spacetime,
	the extension to higher dimensional spacetime
	can easily be done and all of the results we obtained in this paper
	still apply. This is because in $d-$dimensional spacetime,
	massive vector field has $d-1$ degrees of freedom.
	Correspondingly there are two constraints in the theory
	and both of them are of second-class.
	
	In this paper, we have only focused on the dynamics
	in the vector sector. When considering the full,
	a theory whose vector sector passes all the criteria in this paper
	is still not guaranteed to be free from pathologies.
	Although it has three propagating degrees of freedom
	in the vector sector, higher derivatives in background fields
	could potentially introduce ghost degrees of freedom in other sectors.
	For example, the analysis of \cite{Hull:2015uwa}
	hints that for example in the Lagrangian \eq{LgenProca} the counter-terms
	$G_4(Y)R,$ $G_5(Y)G_{\m\n}\na^\m A^\n,$ and $G_6(Y)L^{\m\n\a\b}\na_\m A_\n\na_\a A_\b$
	are required otherwise the gravity sector would possesses ghost degrees of freedom.
	The result in our paper partially confirms this suggestion.
	That is, even without the counter-terms, the Lagrangian \eq{LgenProca}
	should have three propagating degrees of freedom
	as long as the condition \eq{condU0} is satisfied.
	So by limiting ourselves in the vector sector,
	we would not be able to see the pathologies in full theory.
	If the full theory is pathological, then the pathologies
	should be from outside the vector sector.
	
	So an interesting extension of this paper
	is to study dynamics of some other sectors
	as well as the vector one.
	We anticipate that diffeomorphism invariance
	might also help to trivialises many conditions
	to allow us to obtain the required number of degrees of freedom.
	
	An alternative extension would be to relax the condition \eq{condU0},
	but still demand that the Hessian determinant is degenerate.
	In this case, we expect that diffeomorphism invariance
	also trivialises many conditions.

	Diffeomorphism invariance has been of great help
	in degrees of freedom counting in the vector sector
	of generalised Proca theories.
	In fact, we have seen that not all the requirements
	from diffeomorphism invariance are needed.
	In particular, all we need from diffeomorphism invariance
	are to make only eq.\eq{condU0} and
	eq.\eq{TUfull} satisfied.
	Other conditions coming from diffeomorphism invariance requirements
	are not needed.
	So we can still consider the situation where
	diffeomorphism invariance is broken, for example by putting a generalised Proca field
	on a diffeomorphism broken background,
	but eq.\eq{condU0} and eq.\eq{TUfull} are still satisfied.
	It is interesting to explicitly construct these theories.
	
	\subsection*{Acknowledgements}
	We are grateful to Sheng-Lan Ko for interests,
	helpful discussions, and various comments on the manuscript.
	J.S. is supported by an IF Scholarship from the Institute for Fundamental Study
	``The Tah Poe Academia Institute'', Naresuan University,
	and a Research Grant for Graduate Student from the Graduate School, Naresuan University.
	P.V. thanks University of Phayao, and Rajamangala University of Technology Suvarnabhumi
	for hospitality, where part of this work was carried out.
	

\providecommand{\href}[2]{#2}\begingroup\raggedright\endgroup

\end{document}